\documentclass[aps,prb,twocolumn,showpacs,superscriptaddress,10pt]{revtex4-1}

\usepackage{graphicx,amsmath,amssymb,amsfonts,sidecap,color}
\usepackage{epsfig}
\usepackage{epsfig,psfrag,subfigure,amsopn}
\usepackage{graphicx,psfrag,subfigure,float}
\usepackage{xcolor}

\newcommand\ig{\includegraphics}
\newcommand\beq{\begin{equation}}
\newcommand\eeq{\end{equation}}
\newcommand\bea{\begin{eqnarray}}
\newcommand\eea{\end{eqnarray}}

\newcommand\non{\nonumber}

\newcommand\bib{\bibitem}

\makeatletter
\newcommand*{\rom}[1]{\expandafter\@slowromancap\romannumeral #1@}
\makeatother

\begin{document}
\title{Yu-Shiba-Rusinov  States and Ordering of Magnetic\\ Impurities Near the Boundary
 of a Superconducting Nanowire}
\author{Oindrila Deb}
\affiliation{Department of Physics, University of Basel, Klingelbergstrasse 82, CH-4056 Basel, Switzerland}
\affiliation{Department of Physics, University of Surrey, Stag Hill, University Campus, Guildford GU2 7XH, UK }
\author{Silas Hoffman}
\affiliation{Department of Physics, University of Florida, Gainesville, Florida 32611, USA}
\affiliation{Quantum Theory Project, University of Florida, Gainesville, Florida 32611, USA}
\affiliation{Center for Molecular Magnetic Quantum Materials, University of Florida, Gainesville, Florida 32611, USA}
\affiliation{Department of Physics, University of Basel, Klingelbergstrasse 82, CH-4056 Basel, Switzerland}
\author{Daniel Loss}
\author{Jelena Klinovaja}
\affiliation{Department of Physics, University of Basel, Klingelbergstrasse 82, CH-4056 Basel, Switzerland}
\date{\today}

\begin{abstract}
We theoretically study the spectrum induced by one and two magnetic impurities near the boundary of a one-dimensional nanowire in proximity to a conventional $s$-wave superconductor and extract the ground state magnetic configuration. We show that the energies of the subgap states, supported by the magnetic impurities, are strongly affected by the boundary for distances less than the superconducting coherence length. In particular, when the impurity is moved towards the boundary, multiple quantum phase transitions periodically occur in which the parity of the superconducting condensate oscillates between even and odd. We find that the magnetic ground state configuration of two magnetic impurities depends not only on the distance between them but also explicitly on their distance away from the boundary of the nanowire. As a consequence, the magnetic ground state can switch from ferromagnetic to antiferromagnetic while keeping the inter-impurity distance unaltered by simultaneously moving both impurities away from the boundary. The ground state magnetic configuration of two impurities is found analytically in the weak coupling regime and exactly for an arbitrary impurity coupling strength using numerical tight-binding simulations.

\end{abstract}
\maketitle

\section{Introduction}
Magnetic impurities on conventional superconductors exhibit many interesting properties. One such example is the appearance of localized states within the superconducting gap. These states, known as Yu-Shiba-Rusinov (YSR) states,\cite{1,2,3} are induced via the exchange interaction between a magnetic impurity and the superconductor.
The YSR  states have been well studied theoretically~\cite{4,5,6,8,9,10,11,12,14,15,16,17,baur,t4a,tra,meng,pascal,trif,tra,vard,t1}
and observed experimentally in bulk $s$-wave superconductors by scanning  tunneling microscopy (STM) techniques
\cite{7,exp11,13,hatter,ruby,menard,exp4,exp5a,exp5,exp6,exp2,exp3,exp8,exp7} as well as in proximitized semiconducting nanowires with quantum dots by transport measurement techniques \cite{prada, s1,s2,s4,s3,s5}.
Recently, these states have attracted renewed interest in the context of magnetic atomic chains. The YSR states induced by the individual impurities in a magnetic chain can hybridize to form a subgap energy band that can host Majorana bound state (MBS)~\cite {23,18,24,25,kotetes,19,20,21,22,sh,ando,awoga,thriler,black,rev2,rev3,t4}. The zero-energy bias peaks has been recently observed in such chains\cite{26,r1,27,feld,w1}. However, the formation of such MBSs critically depends on the magnetic order  inside the spin chain. This magnetic order is itself determined by the effective exchange interaction between the impurities that is mediated by the underlying superconductor.

\begin{figure}[!t]
\vspace{1.5cm}
\includegraphics[width=8.5cm]{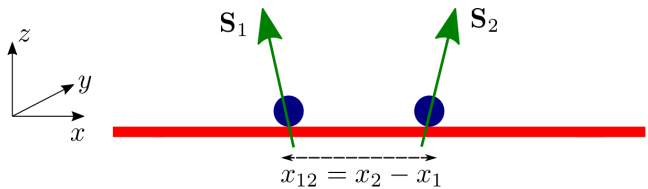}
\hspace{-.5cm}
\caption{One-dimensional nanowire with superconducting gap aligned along  the $x$ axis with two magnetic impurities with classical spins ${\bf S}_1$ and ${\bf S}_2$ separated by the inter-impurity distance $x_{12}= x_2-x_1$. The impurity spins are exchange coupled to the quasiparticles in the nanowire. This gives rise to YSR subgap bound states (not shown) and effective interactions between the magnetic impurities which are both affected by the boundaries of the nanowire.
}
\label{fig01}
\end{figure}

When the exchange interaction is small compared to the Fermi energy, the effective interaction between two magnetic impurities (see Fig. \ref{fig01}) in such a system is mediated via the quasiparticles in the superconductor and, is well-described by the Ruderman-Kittel-Kasuya-Yosida (RKKY) interaction \cite{28a,28b,28c,29,30,31,32,sch,bruno,bruno2,egger,schaffer,chesi,giulani,kogan,hsu1,hsu2,klinovaja,t6,t7,t8,t9,t10,t5,flensberg,henry}. 
The RKKY interaction between two spin impurities located inside the bulk of the system results in the magnetic ordering of the Heisenberg-type in the absence of  spin-orbit interaction and depends only on the relative angle between the impurity spins, ensuring that the ground state magnetic configuration is either ferromagnetic (FM) or antiferromagnetic (AFM). As the sign of the effective exchange interaction oscillates as a function of the inter-impurity distance, the magnetic ground state, likewise, oscillates between FM and AFM ordering. When the exchange interaction between the impurity and quasiparticles is increased beyond the Fermi energy of the superconductor, the approximations invoked by the RKKY interaction break down and the the ground state of the magnetic impurities departs from such a simple description. This is because (1) the coupling of the impurities to the quasiparticles can no longer be treated perturbatively (as known from gapless systems~\cite{flensberg})  and (2) the YSR states can be close to the chemical potential and thereby strongly renormalize the superconducting gap under the impurity~\cite{17}.

However, a description of magnetic impurities close to the boundaries of realistic finite-size samples received very little attention so far. Studies of this type are particularly relevant for low-dimensional systems, where the superconductivity is induced via the proximity effect by a bulk superconductor, because both the longer range of the RKKY interaction and reduction or absence of a power-law decay of the YSR wavefunctions.
Motivated by this, in this work we study how the boundary of such proximitized superconducting systems modifies the energy of the YSR states and, subsequently, the magnetic ground state.  In the following, we consider one and two magnetic impurities placed close to the boundary of an effective semi-infinite one-dimensional (1D) superconductor, see Fig. \ref{fig01}.
Such a 1D system is particularly suitable for our analysis as (1) we are able to obtain analytic results for the  YSR energies and the RKKY interaction and (2) we expect an enhancement of the boundary effects in such 1D set-ups as compared to magnetic impurities embedded in two- and three-dimensional superconductors~\cite{st}.

For one magnetic impurity, we find a position-dependent energy of the induced YSR bound state. Consequently, for sufficiently strong exchange interaction, the superconducting condensate undergoes multiple quantum phase transitions as the distance to the boundary is changed. For two magnetic impurities, we find phase transitions between the FM and AFM ground state by changing only the distance to the boundary and keeping the inter-impurity distance fixed.  In the weak exchange interaction limit, we analytically show that by tuning the inter-impurity distance appropriately, the effect of the boundary can occur even when the impurities are deep in the bulk of the nanowire. When the exchange interaction between the impurity and quasiparticles is large and cannot be treated within the RKKY framework, we  find the magnetic ground state numerically and again observe a similar dependence of the ground state configuration on the distance to the boundary. In this limit, the contributions from both the quasiparticle states and the YSR bound states become crucial in determining the ground state of the system.

The outline of the paper is as follows. In Sec.~\ref{analytic}, we introduce  a continuum Hamiltonian  of an effective 1D superconductor   hosting the magnetic impurities and find the Green's function in the presence of the boundary analytically. We first describe the analytical results for the YSR energy of a single magnetic impurity in a semi-infinite superconductor in the Sec. \ref{oneimp}. This is followed by a study of the RKKY interaction between two magnetic impurities in a semi-infinite 1D superconducting wire in  Sec.~\ref{twoimp}. Next, using a discretized Hamiltonian corresponding to the continuum model, we numerically include contributions to the total energies of both the subgap YSR states and the supragap quasiparticle states to determine the ground state configuration in Sec.~\ref{numeric}.  Though the ground state configuration is primarily determined by the contribution from the bulk states, we show that the YSR states dominate in determining the phase boundary between the FM and the AFM phases in the limit of large exchange interaction. Finally, we present a summary of our results.

\section{Analytical Results}
\label{analytic}

We consider two classical magnetic impurities placed on a 1D wire, aligned along the $x$-axis, in proximity to an $s$-wave superconducting (SC) substrate, see Fig.~\ref{fig01}. The impurities $\textbf{S}_1$ and $\textbf{S}_2$ are located at a distance $x_1$ and $x_2$ from the boundary of the wire, respectively. The inter-impurity distance is denoted by the relative coordinate $x_{12} = x_2-x_1$.   

The system is described by the Hamiltonian, $H=H_0+H_{\textrm{imp}}$, which is a sum of the kinetic term, of the superconducting pairing term, and of the exchange term describing coupling between the spins of  magnetic impurities to the electrons in the 1D wire,
\begin{align}
 &H_0 = \frac{1}{2}\int d{x} ~~ \Psi^{\dagger}(x)\biggl[\left(-\frac{\hbar^2}{2m}\frac{d^2}{dx^2}-\mu\right)\tau_z+\Delta\tau_x \biggr]\Psi(x) \non \\
&\hspace{70pt}\equiv\frac{1}{2}\int d{x} ~~\Psi^{\dagger}(x)\mathcal H_0\Psi(x)\,, \label{eq:hcont1}\\
&H_{\textrm{imp}}=\frac{JS}{2} \sum_{n=1,2}  \Psi^{\dagger}(x_n)~{{\bf s}_n}\cdot {\boldsymbol \sigma}\Psi(x_n)\non \\
&\hspace{70pt}\equiv\frac{1}{2}\sum_{n=1,2}\Psi^{\dagger}(x_n)\mathcal H^{(n)}_\textrm{imp}\Psi(x_n)\,, \label{eq:hcont2}
\end{align}
respectively, and $\mathcal H_0$ and  $\mathcal H^{(n)}_\textrm{imp}$ are referring to the corresponding Hamiltonian densities. The Pauli matrices $\boldsymbol {\sigma}_{x,y,z}$ ($\boldsymbol{\tau}_{x,y,z}$) operate in spin (Nambu) space. The Hamiltonian is written in a basis which corresponds to the four-component Nambu operator $\Psi(x) = [\psi_{\uparrow},\psi_{\downarrow}, \psi^{\dagger}_{\downarrow},-\psi^{\dagger}_{\uparrow}]^T$, where $\psi_{\sigma}(x)$  is the electron field operator with spin $\sigma=\uparrow,\downarrow$. Here, $\mu$ denotes the chemical potential, $\Delta$ is the  superconducting pairing strength (induced by the proximity effect) and $J$ denotes the strength of the exchange coupling between magnetic impurities and the electrons in the superconducting wire. We assume $J>0$ without loss of generality such that the exchange interaction is antiferromagnetic. The magnitude $S$ of the impurity spin is much larger than unity so that quantum spin-fluctuations are negligible and, therefore, $\textbf{S}$  is treated as a fixed classical spin vector. Although we are going to  focus on the case of identical magnetic impurities, which is substantially simplifying our analytical expressions, the directions of the magnetic impurities, $\textbf{s}_n=(\sin{\theta_n}\cos{\phi_n},\sin{\theta_n}\sin{\phi_n},\cos{\theta_n})$, can be different. We also note that, due to the spin rotation symmetry of the system, the magnetic ground state depends only on the relative angle between two impurity spins. It is a straightforward task to generalize our model to treat magnetic impurities of different strengths.

The full Green's function, $G=(E+i0^+-\mathcal H)^{-1}$, corresponding to the energy $E$, where $i0+$ represents an infinitesimal small imaginary shift in energy, is written  in position representation as $G(x_1,x_2;E)= \langle x_1|(E+i0^+-\mathcal H)^{-1}|x_2\rangle$.

In the absence of impurities, the unperturbed Green's function $G_0$ is obtained by replacing 
$\mathcal H$ by $\mathcal H_0$. For a translational invariant 1D SC, i.e. in the absence of boundaries, we find $G_0(x_2,x_1;E)\equiv G_0(x_{12};E)$, with

\begin{align}
G_0(x;E)=- i \pi \nu_F \Big[ &\frac{E+\Delta\tau_x }{\sqrt{E^2-\Delta^2}} \cos(k_Fx)  \non \\
+&i \tau_z \sin(k_F|x|) \Big] \,e^{-|x|/\xi_E},
\label{singG} 
\end{align}  
where we have assumed that the energy $E$ is counted from the chemical potential and $|E|\leq \Delta$. Here, $\xi_E=\sqrt{\Delta^2-E^2}/\hbar v_F$ is the exponential decay length due to the gap, $v_F$ the Fermi velocity, and $\nu_{F}=m/(\pi\hbar^2k_F)$ the 1D density of states (per spin) at the Fermi energy of the metallic (gapless) phase, with $k_F$ the Fermi wavevector. Below, we will also need the Green's function for energies $|E|\gg \Delta$. In this case, we can just use the Green's function for gapless systems

\bea
G_0^0(x;E)=-i \frac{m}{{\hbar}^2 (k+i0^+)} e^{ik|x|},
\label{singG0}
\eea  
where $k= \sqrt{2m(E+\mu)}$. 
We note that in 1D there is no  power-law decay prefactor in terms of $|x|$   in above Green's function (with and without gap) in contrast to higher dimensions~\cite{meng}.

In a semi-infinite 1D wire with a boundary at $x=0$, the wave function $\psi_{b,\sigma}(x)$ must satisfy  vanishing boundary conditions at $x=0$. Hence, 
\begin{align}
\psi_{b,\sigma}(x)=\frac{1}{\sqrt{2}}[\psi_{\sigma}(x)-\psi_{\sigma}(-x)]\,,
\end{align}
where $\psi_\sigma(x)$ is the wave function in the bulk. 
Consequently, using the eigenstates, $\mathcal H_0 \psi_n=E_n\psi_n$, for the Green's function representation, 
$G_0(x_1,x_2;E)= \sum_n  (E+i0^+- E_n)^{-1} \psi_n^*(x_1) \psi_n(x_2)$, we see that the Green's function for the semi-infinite system has the corresponding form,
 \begin{align}
 G_{b}(x_1,x_2;E)= G_{0} (x_{12};E)-G_0 (X;E),
 \label{gb2}
\end{align}
where $X=x_2+x_1$. We note that  the Green's functions are diagonal in spin-space as the Hamiltonian  $\mathcal H_0$ is spin-independent.
 
\subsection{Single Magnetic Impurity}
\label{oneimp}
We first explore the energy of the YSR state induced by a single magnetic impurity in the vicinity of the boundary. In this case, as there is no contribution from the RKKY interaction, we focus only on the renormalization of the YSR energy. In the standard case of a magnetic impurity located deeply inside the bulk of a 1D system, the energy of the YSR bound states is well-known~\cite{1,2,3} and  given by $\pm \bar E_{s}$, where $\bar E_s=~\Delta(1-\alpha^2)/(1+\alpha^2)$, with $\alpha=\pi \nu_{F} J S$.

In the presence of a boundary, we start from the Dyson equation~\cite{meng}, $G=G_0+G_0 \mathcal V G$, with $\mathcal V =\sum_{n=1,2}
\mathcal H^{(n)}_\textrm{imp} \delta (x-x_n)$. Taking position-state matrix elements of this equation, and keeping only one impurity at a distance $x_1$ to the boundary of the 1D SC, we find $G(x_1,x;E)(1-G_0(x_1,x_1;E)\mathcal H^{(1)}_\textrm{imp})=G_0(x_1,x;E)$. For vanishing boundary conditions,  $G_0(x_1,x_1;E)$ is given by $G_b(x_1,x_1;E)$.
The subgap bound state energy $E_s$ is then found from the pole of  $G(x_1,x;E)$ and thus must satisfy the equation $\det[1 -G_b(x_1,x_1;E_s) \mathcal H_\textrm{imp}^{(1)}]=0$.

Assuming the coherence length being weakly dependent on the YSR energy, i.e. $\hbar v_F/\sqrt{\Delta^2-E^2}\rightarrow\hbar v_F/\Delta$ in the exponent of Eq.~(\ref{singG}), we obtain the YSR bound state energy $E_s$ after a straightforward calculation:

\begin{widetext}
\begin{align}
&\frac{E_s}{\Delta}=\frac{1-\alpha^2+2\alpha^2\cos(2k_F x_1) A -\alpha^2 A^2}
{\sqrt{
(1+ \alpha^2)^2 + 4 \alpha^4 A^2 + A^4
       \alpha^4 - 
       4 \alpha^2 A [1 +  (1+ A^2) \alpha^2] \cos(2k_F x_1) + 
       2 A^2 \alpha^2 (1 + \alpha^2) \cos(4 k_F x_1)}}
       \label{full},
\end{align}
\end{widetext}
where $A=e^{-2x_1/\xi_{sc}}$ and $\xi_{sc}=\hbar v_F/\Delta$ is the superconducting coherence length. 

For the impurity sufficiently away from the boundary, $x_1\gg \xi_{sc}$, this expression considerably simplifies, reducing to the bulk expression for the bound state, 
$\bar E_{s}$, 
with an exponentially small correction that exhibits Friedel oscillations induced by the boundary:

\begin{align}
\frac{E_s}{\Delta} \approx \frac{1-\alpha^2}{1+\alpha^2} +  \frac{4\alpha^2}{(1+\alpha^2)^2} \cos(2k_F x_1) e^{-2x_1/\xi_{sc}}.
\label{fullapprox}
\end{align}

\begin{figure}[t]
\includegraphics[width=\linewidth]{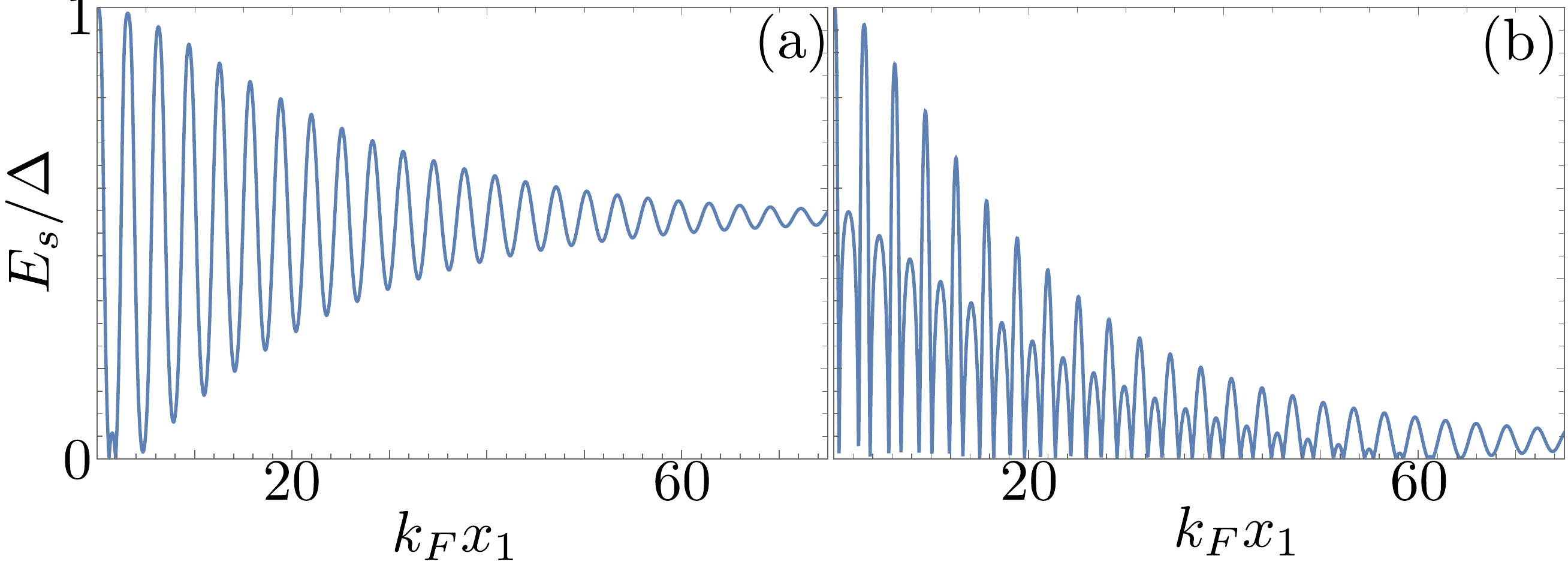}
\caption{  The  energy of the YSR state $E_s/\Delta$ [see Eq.~(\ref{full})] as a function of the distance $k_F x_1$ of the impurity away from the boundary for (a) $\alpha=0.55$  and (b) $\alpha=0.96$. The energy $E_s$ oscillates around its bulk value. (b) If the YSR energy is close to zero energy for a magnetic impurity placed far away from the boundary, then moving it closer to the boundary can induce a quantum phase transition and change the occupancy of the YSR bound state.
The coherence length is fixed to $k_F\xi_{sc}=40$.} 
\label{analytic1}
\end{figure}

In Fig.~\ref{analytic1}, showing only the positive energy solutions, we consider two typical cases:  $\alpha \ll 1$ and  $\alpha \approx 1$.  If the impurity is relatively weak, $\alpha \ll 1$, the YSR state energy is inside the superconducting gap and still away from zero energy if the impurity is placed far away from the boundary, see Fig.~\ref{analytic1}(a). Close to the boundary, the YSR energy oscillates around the bulk value $\bar E_{s}$ with the period given by $\lambda_F/2$, where $\lambda_F=2\pi/k_F$ is the Fermi wavelength. If the impurity is stronger, $\alpha \lesssim 1$, the YSR energy is close to but still above zero energy for impurities far away from the boundary, see  Fig.~\ref{analytic1}(b). Moving the impurity closer to the boundary drives the YSR state to a negative energy [due to the second term in Eq.~(\ref{fullapprox})] which induces a quantum phase transition of the ground state from even parity to odd parity~\cite{8,9,10,11,12,17}. As the energy continues to oscillate between positive and negative values as the impurity approaches the boundary, the system undergoes a series of quantum phase transitions in which the parity of the ground state oscillates. 
The parity change occurs at positions $x_1$ for which $E_s(x_1)=0$. From Eq.~(\ref{full}) we see that this is the case if $x_1>0$ satisfies  the transcendental
equation
\begin{align}
\cos(2k_F x_1) =\frac{1}{2} e^{-2x_1/\xi_{sc}} - \frac{1 -\alpha^2 }{2\alpha^2} e^{2x_1/\xi_{sc}}.
\label{eq2}
\end{align}

From this expression we conclude that the quantum phase transitions are possible for values of $\alpha$  close to one. Generally, Eq. (\ref{eq2}) has multiple solutions,  see Fig. \ref{analytic1}(b). 
 Evidently, the parity of the ground state can be chosen by appropriate positioning of a magnetic impurity with respect to the boundary. We note that in this work we have neglected the local effect of the magnetic impurity on the superconducting order parameter, which should be determined self-consistently~\cite{12,meng}. In this case, the YSR energy will change discontinuously at zero energy.

\subsection{Two Magnetic Impurities}
\label{twoimp}

Next, we consider two identical magnetic impurities located at positions $x_1$ and $x_2$, respectively. There are now two energetic contributions which we need to consider: One are the YSR energies
of the bound states associated with each impurity, 
and the other one is  the RKKY interaction between the two magnetic impurities transmitted by the electrons of the superconductor.
First, the energy spectrum, generally, contains two in-gap YSR states. Similarly to the single impurity case, the two-impurity YSR energies can be found by determining the poles of the Green's function dressed by scattering from two impurities. Using again the Dyson equation, we readily find that the energies must satisfy the following equation:
\begin{widetext}
\begin{equation}
\det\left\{1-[1-\mathcal H^{(1)}_\textrm{imp}G_b(x_1, x_1;E_s)]^{-1} \mathcal H^{(2)}_\textrm{imp}G_b(x_1,x_2;E_s)[1-H^{(2)}_\textrm{imp}G_b(x_2, x_2;E_s)]^{-1}H^{(1)}_\textrm{imp}G_b(x_2,x_1;E_s)\right\}=0.
\end{equation}

\end{widetext}
Although far away from the boundary, $x_1,x_2\gg\xi_{sc}$, the solution of this equation can be found analytically~\cite{17}, the analysis is considerably more complicated when the impurities are near the boundary of the wire. We thus postpone a discussion of this case to Sec.~\ref{numeric} where we solve the problem exactly within a tight-binding approach.

Second, we turn now to the RKKY interaction, which is valid when  the exchange coupling $J$ is weak, and $H_\textrm{imp}$ can be treated perturbatively.  In this case, the YSR states are near the gap edge and their energies and correlations can be neglected.

Following the usual RKKY type of analysis~\cite{28a,28b,28c,29,30,31,32,sch,bruno,bruno2,egger,schaffer,chesi,giulani,kogan,hsu1,hsu2,klinovaja,t6,t7,t8,t9,t10,t5}, we find an effective exchange interaction between magnetic impurities ${\textbf s}_{i} $ located at positions $x_i$, $i=1,2$, given by the following expression:
\begin{align}
&H^{\textrm{RKKY}} _{1,2}=-\frac{(JS)^2}{\pi} \textrm{Im} \int^{E_{F}}_{-\infty} dE ~ \textrm{Tr}\bigl[({\textbf s}_1 \cdot {\boldsymbol \sigma})\non\\
&\times G_b({x_1, x_2};E) ({\textbf s}_2 \cdot {\boldsymbol \sigma}) G_b({ x_2,  x_1};E)\bigr],
\label{eq:rkky}
\end{align} 

where $E_F$ is the Fermi energy and $\textrm{Tr}$ is the trace over the electron spin degrees of freedom. 
When the distance between the impurities is smaller than $\xi_{sc}$, the superconducting correlations can be  neglected and, using Eq.~(\ref{gb2}) with Eq.~(\ref{singG0}) we find that Eq.~(\ref{eq:rkky}) reduces to
\begin{align}
H^{\textrm{RKKY}} _{1,2}&=\frac{2m(JS)^2}{\pi\hbar^2} \biggl[F(x_{12})+F(X)-2F \left(x_2\right)\biggr] {\textbf s}_1 \cdot {\textbf s}_2,\non \\
&\equiv \frac{2m(JS)^2}{\pi\hbar^2}  ~F_{\textrm{sum}} (x_1,x_2)~{\textbf s}_1 \cdot {\textbf s}_2,
\label{eq:hbrkky}
\end{align} 

where $F(x)= Si(2k_{F} x)-\pi/2$, with $Si(y)$ denoting the sine integral function. $\lambda_F=2\pi/k_F$ is the Fermi wavelength and  
The asymmetric dependence of the RKKY Hamiltonian on $x_1$ and $x_2$ originates from our assumption $x_2\geq x_1$, i.e. $x_{12}\geq 0$. Evidently, the interaction between the impurity spins is of Heisenberg type. The ground state configuration is ferromagnetic for $F_{\textrm{sum}}<0$ and antiferromagnetic  for $F_{\textrm{sum}}>0$. Although $F{(x)}$ oscillates, upon averaging over a Fermi wavelength, we expect $F(x_{12})$ to be generally a dominant term in Eq.~(\ref{eq:hbrkky}) as $F(x)$ scales inversely with the distance $x$ and $x_{12}\leq x_2\leq X$. For distances greater than the coherence length $\xi_{sc}$, the RKKY interaction is exponentially suppressed. When the impurities are located far away from the boundary while remaining close to each other such that $x_1, x_2 \gg \xi_{sc} \gg x_{12}$, $F(X)$ and $F(x_2)$ can be neglected  and Eq.~\eqref{eq:hbrkky} takes the usual form of the RKKY interaction in the absence of the boundary effects \cite{28a,28b,28c,bruno,sch,bruno2,36,37} with the position dependence given by $ H^{\textrm{RKKY}} _{1,2}\propto F({x_{12}})~{\textbf s}_1 \cdot {\textbf s}_2$.

\begin{center}
\begin{figure}
\hspace{-2.5cm}
\includegraphics[width=\linewidth]{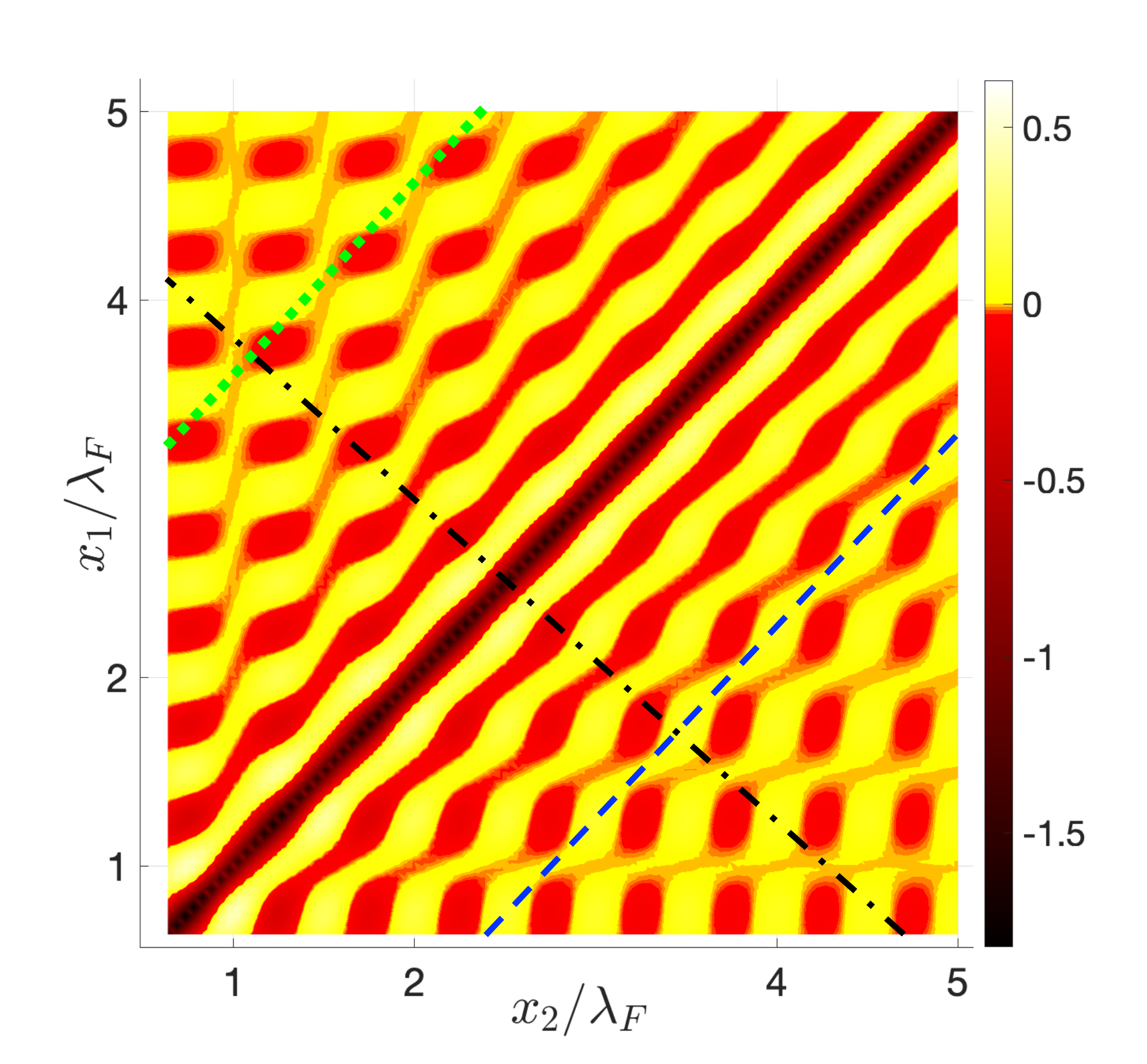}
\hspace{-2cm}
\caption{Colour plot of the RKKY coefficient $F_{\textrm{sum}}$ as a function of $x_{2}/\lambda_F$ on the $x$-axis and $x_1/\lambda_F$ on the $y$-axis, where $\lambda_F=2\pi/k_F$ is the Fermi wave-length. The red (yellow) color denotes the region in which $F_{\textrm{sum}}$ is negative (positive) and hence the ground state is FM (AFM).
The green dotted line and the blue dashed line parallel to the diagonal are two lines of constant $x_{12}$, while $X$ remains constant along the black dash-dotted line perpendicular to them.The figure exhibits oscillations between the FM and AFM ground state along the constant $x_{12}$ lines, unlike the conventional RKKY interaction obtained far away from boundaries.}
\label{fig:rkky}
\end{figure}
\end{center}

In Fig.~\ref{fig:rkky}, we plot $F_{\textrm{sum}}(x_1,x_2)$ as a function of  $x_{1}$ and $x_{2}$, where the red (yellow) regions denote $F_\textrm{sum}<0$ ($F_\textrm{sum}>0$) indicating a FM (AFM) ground state.   The lines parallel to the diagonal $x_2=x_1$, i.e. $x_{12}=0$, are the regions of constant $x_{12}$, while $X$ remains constant along the lines parallel to the anti-diagonal, $x_2=-x_1$. In the region around the line $x_{12}=0$, the RKKY coefficient $F_{\textrm{sum}}$ does not change sign along the constant $x_{12}$ lines, implying no transition in the ground state spin configuration as $x_{12}$ is unaltered. This arises from the fact that for $x_{12}\rightarrow 0$, the RKKY Hamiltonian $H^{\textrm{RKKY}} _{1,2}$ defined in Eq. \eqref{eq:hbrkky} is dominated by $F(x_{12})$. Therefore, near these points in parameter space, the ground state configuration is only a function of $x_{12}$, similar to the conventional RKKY interaction in the absence of boundaries. However, with increasing magnitude of $x_{12}$, $|F(x_{12})|$ decreases and even goes through zero and, as a consequence, the two other terms $F(X)$ and $F(x_2)$ in Eq.~\eqref{eq:hbrkky}  become significant. The interplay between these terms can then induce oscillations in $F_{\textrm{sum}}$, even along lines of constant $x_{12}$. In particular, for large $x_{12}$, e.g. $|x_{12}|=2.6\lambda_F$ (green dotted line in Fig.~\ref{fig:rkky}), the transitions between the FM and the AFM phases upon changing $X$ are particularly pronounced. Evidently, these transitions in the magnetic ground state are solely due to the boundary effects whose contributions are encoded in the terms $F(X)$ and $F(x_2)$ in the RKKY Hamiltonian. Conversely, for some values of $x_{12}$, e.g. the blue dotted line in which $|x_{12}|=1.75\lambda_F$, the ground state configuration is independent of $X$. We always observe oscillations between the FM and the AFM phases along the lines of constant $X$ (black dash-dotted line in Fig.~\ref{fig:rkky}) as $F(x_{12})$ generally dominates over $F(X)$.

Qualitatively, one can easily interpret Fig.~\ref{fig:rkky} in the limit $x_1$, $x_2$, $x_{12}\gg \lambda_F$, in which $Si(2k_F x)-\pi/2 \approx \cos(2  k_Fx)/2  k_Fx $, wherein we can make use of the following approximation \cite{bruno} 
\begin{align}
\hspace{-.4cm}
F_\textrm{sum}&\approx \frac{\cos(2 k_F x_{12})}{2 k_Fx_{12}}+\frac{\cos(2 k_FX)}{2 k_FX}-\frac{\cos(2 k_F x_2)}{k_F x_2}.
\label{eq:coeff}
\end{align} 
If the inter-impurity distance $x_{12}$ is chosen such that $ 2 k_F x_{12}  \approx (2n+1)\pi/2$, where $n$ is an integer, then $F(x_{12}) \approx 0$. Thus, at such values of the inter-impurity distances, the RKKY coefficient $F_{\textrm{sum}}$ will be dominated by the boundary induced terms $F(X)$ and $F(x_2)$. The green dotted line in Fig.~\ref{fig:rkky} corresponds to $|x_{12}|= 2.6 \lambda_F$ ($n \approx 10$). Hence, as $x_1$ and $x_2$  are modified keeping $x_{12}$ unaltered, oscillations in the ground state configuration can be seen along this line, originating from the interplay between $F(X)$ and $F(x_2)$ in Eq. \eqref{eq:coeff}. On the other hand, when $2 k_F x_{12}  \approx n \pi$, $|F(x_{12})|$ is a local maximum and hence dominates over the other two terms in Eq.~\eqref{eq:coeff}. The blue dashed line with $|x_{12}|=1.75 \lambda_F$  in  Fig.~\ref{fig:rkky} satisfies this condition ($n\approx 7$) and the ground state exhibits the predicted behaviour.

\section{Numerical results}
\label{numeric}

In this section we depart from an analytical analysis and use a tight-binding Hamiltonian description of our two impurity system. This allows us to go beyond the small $J$ limit to find numerically  the energies of hybridized YSR states and to determine the magnetic ground state of the two impurities for any value of $J$.

The tight-binding Hamiltonian has the following form:
\begin{align}
H_0=-t &\sum_{n,\sigma} c_{n,\sigma}^\dagger c_{{n+1},\sigma} -\frac{\mu_l}{2} \sum_{n,\sigma} c_{n,\sigma}^\dagger c_{ n,\sigma}\non\\&+\sum_{n} \Delta c^{\dagger}_{{n},\uparrow} c^{\dagger}_{{n},\downarrow}+ H.c.,\non\\
H_{\textrm{imp}}=& JS\sum_{n} c^{\dagger}_{n} {\textbf s}_n \cdot {\boldsymbol \sigma}(\delta_{n,n_1}+\delta_{n,n_2})c_n .\label{eq:h0}
\end{align}
where $c_n=\left[ {\begin{array}{cc} c_{n,\uparrow}, & c_{n,\downarrow} \end{array}} \right]^T$ and $c_{n,\sigma}$ is the annihilation operator acting on an electron with spin $\sigma=\uparrow,\downarrow$ at a lattice site $n=x/a$, $a$ being the lattice spacing; $t$ is the hopping amplitude, $\mu_l$ denotes the chemical potential and $J$ denotes the exchange interaction strength between the impurity and the substrate. We have identical magnetic impurities, while the spin directions $\textbf{s}_n=(\sin{\theta_n}\cos{\phi_n},\sin{\theta_n}\sin{\phi_n},\cos{\theta_n})$ of the magnetic impurities can be different, as also considered previously. The spin-rotation symmetry of the system ensures that the magnetic ground state depends only on the relative angle $\theta=\theta_2-\theta_1$ between the impurity spins. The total number of lattice sites is $N$. We define a quantity $\tilde{{J}}=JS$ which we will use later in our calculation to simplify the representation of results. As usual, the tight-binding description is just the discretized version of the continuum Hamiltonian $H$ given in Eq.~(\ref{eq:hcont2}), and as such the lattice spacing $a$ has no relation to an atomistic structure, it is just a  discrete length chosen sufficiently small such that the numerics converges and the tight-binding description becomes an accurate approximation of the continuum model.

\subsection{Single Magnetic Impurity}

\begin{figure}[t]
\begin{center}
\subfigure[]{\ig[width=0.48\columnwidth,keepaspectratio]{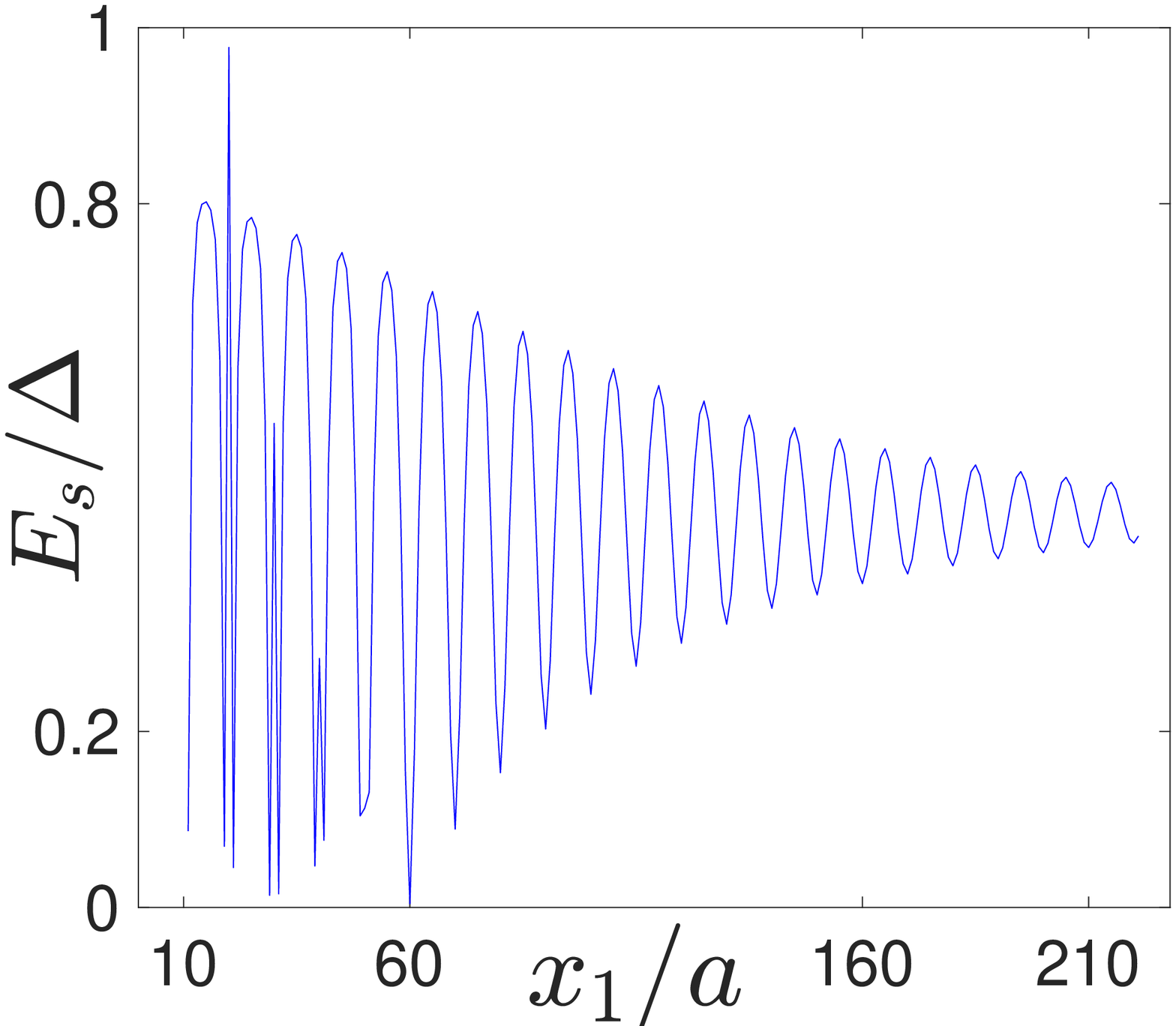}}
\subfigure[]{\ig[width=0.48\columnwidth,keepaspectratio]{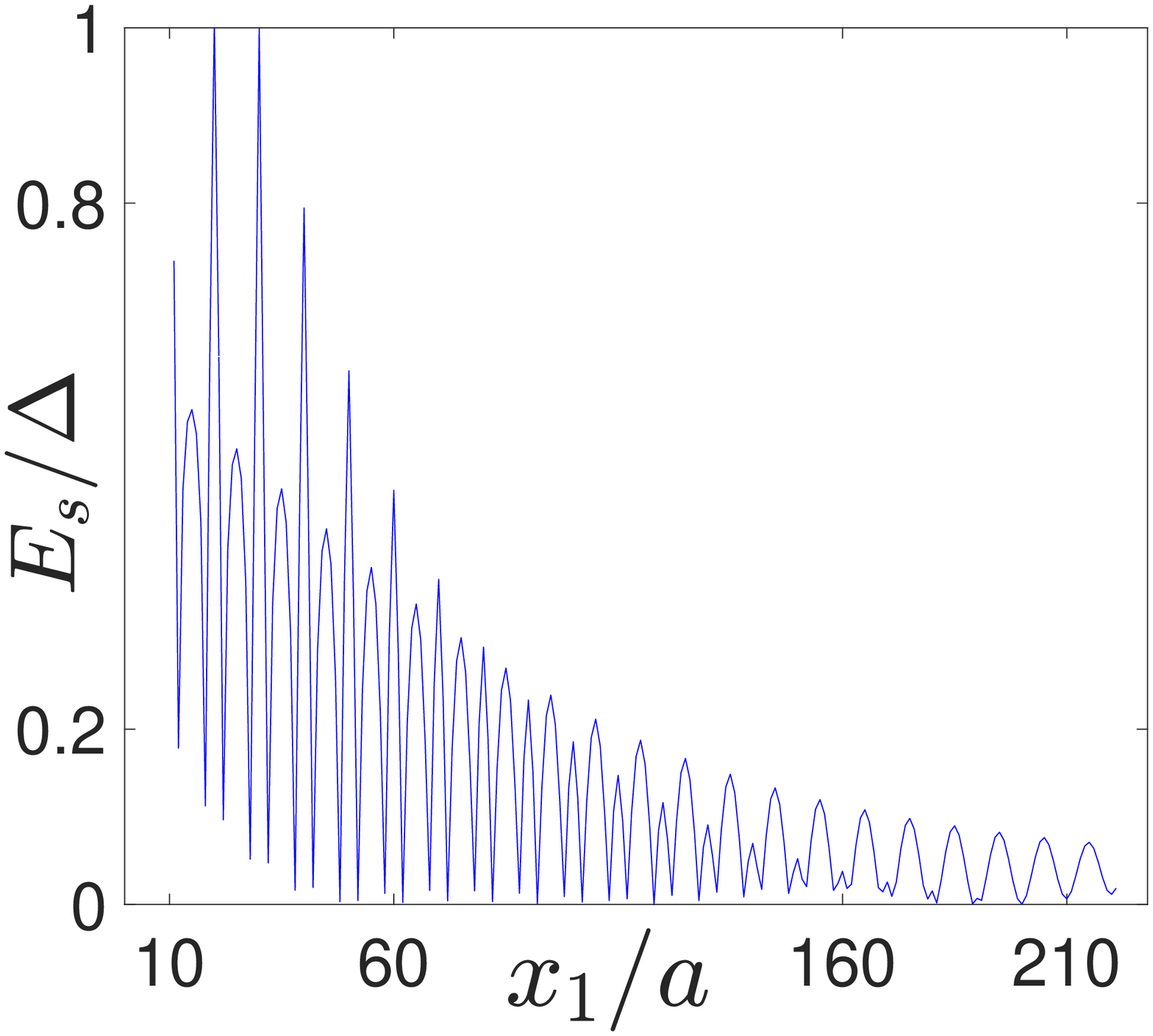}} 
\end{center}
\caption{The figure shows YSR energy $E_s/\Delta$ as a function of the distance $x_1/a$ of the impurity away from the boundary for (a)  $\tilde{J}=0.644 t$   and (b) $\tilde{J}=1.0 t$. We used the following values of the parameters: $N=1000$, $\mu_l=-1.9t$, $\Delta=0.005t$. As the impurity approaches the boundary, the YSR energy oscillates with a periodicity of $\lambda_F/2$ around its value obtained deep inside the bulk. The amplitude of oscillations decay exponentially due to the presence of the  superconducting gap. In Fig. (b), close to the boundary, the YSR energy crosses the chemical potential indicating the quantum phase transition.}

\label{imp1}
\end{figure}

 We first check our results for a single impurity discussed in Sec.~\ref{oneimp} using the above tight-binding Hamiltonian. The energy of the YSR bound states are calculated numerically by diagonalizing the lattice Hamiltonian and plotted as a function of the dimensionless distance $x_1/a$ between the impurity and the boundary. We choose $\mu_l=-1.9 t$ and $\Delta=0.005t$,  which corresponds to a Fermi wave length $\lambda_F\approx 20a$ and a superconducting coherence length $\xi_{sc} \approx 126 a$, respectively. For such a small value of the chemical potential, $\lambda_F\gg a$, the deviation from the quadratic dispersion due to the higher-order terms is small.  Hence, in this limit, the lattice model provides  a good description of the continuum model. In Fig.~\ref{imp1}(a) and (b), we plot the YSR bound state energy as a function of distance $x_1$ awy from the boundary for  $\tilde{J}=t$ and $\tilde{J}=0.644t$, respectively. We choose our parameters such that for impurities far away from the boundary the energy of the corresponding YSR states are the same as in Fig.~\ref{analytic1} for the continuum case.  As the impurity is moved closer to the boundary, we observe oscillations in the YSR energies around their bulk values. The oscillations have a periodicity of $\lambda_F/2$ with an exponentially decaying amplitude of the form $e^{-x_1/\xi_{sc}}$ originating from the superconducting gap, in accordance with the analytical result. In Fig.~\ref{imp1}(b), the YSR energy lies close to zero when the impurity in placed deeply inside the bulk of the system and oscillates around zero as the impurity is placed closer to the boundary. Evidently, this indicates the quantum phase transitions, in which the parity of the ground state of the system changes as the impurity-boundary distance is altered. These features, observed numerically, are in very good agreement with the analytical results for the YSR energies presented in Fig.~\ref{analytic1}.

\subsection{Hybridization between the YSR states} 
\label{hyb}
\begin{figure}[tb]
\begin{center}
\subfigure[]{\ig[width=0.475\columnwidth]{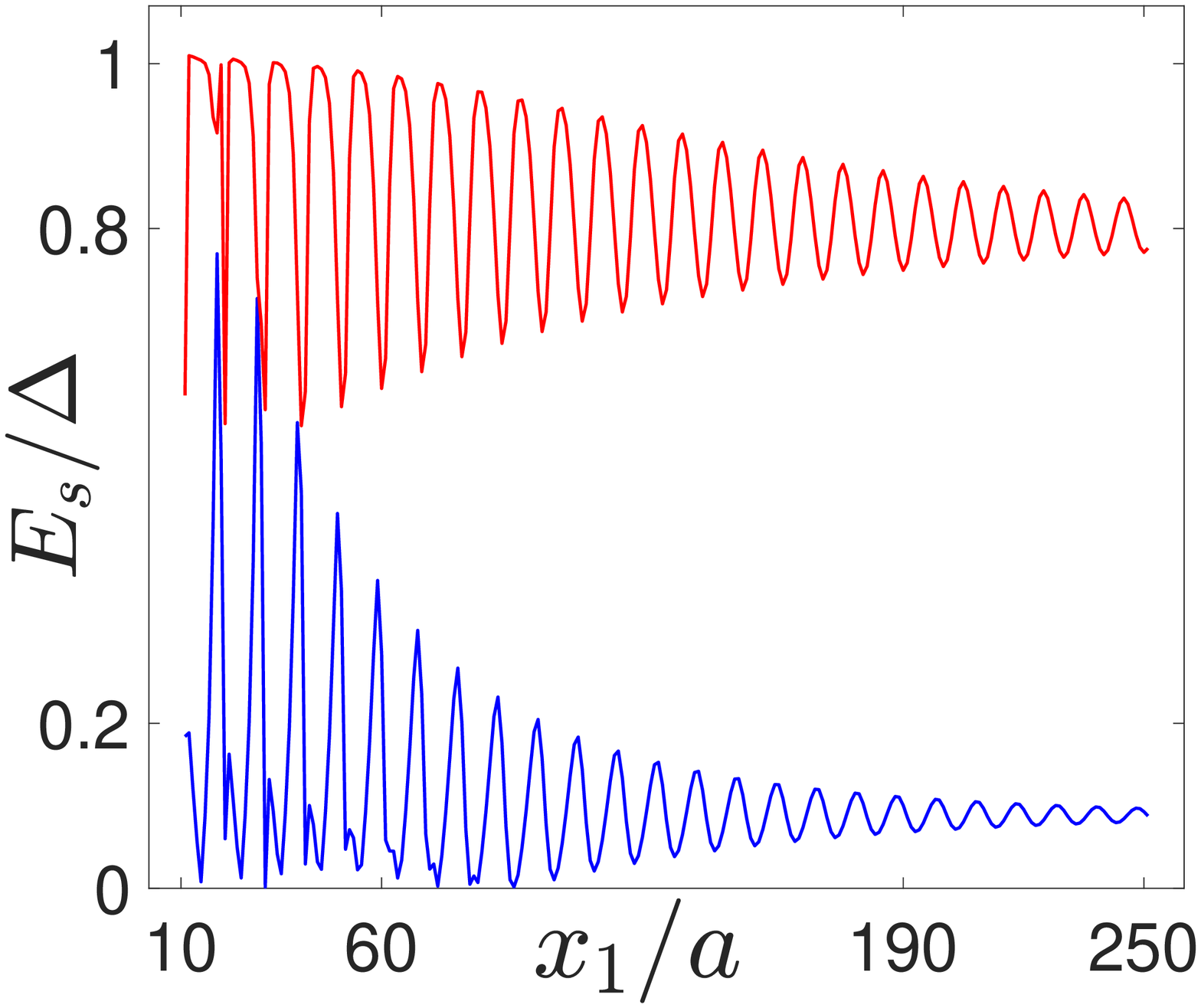}}
\subfigure[]{\ig[width=0.475\columnwidth]{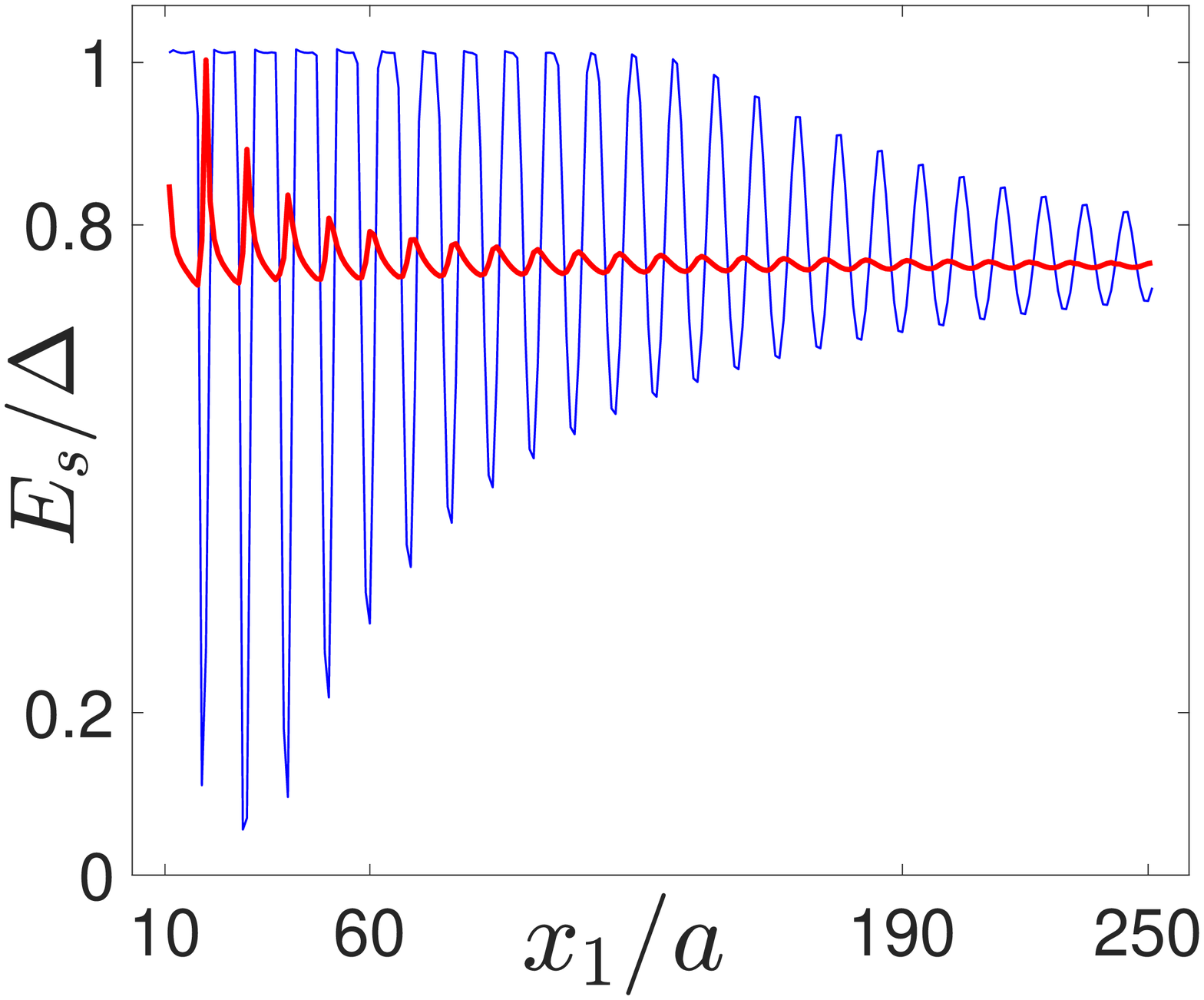}}
\end{center}
\caption{The energies $E_s$ for two YSR states (blue and red)
for (a) the FM configuration and (b) the AFM configuration of the magnetic impurities as a function of the distance $x_1/a$ between the first impurity $\textbf{s}_1$ and the boundary for a fixed inter-impurity distance $x_{12}=28 a$. Note that in (a) the two YSR states are strongly hybridized whereas this is not the case in (b). In (a) the hybridized energy levels of the YSR states oscillate around their bulk value with an increasing amplitude of oscillation close to the boundary. The YSR energies for the AFM configuration in (b) do not hybridize with each other but exhibit oscillations induced by the boundary. Each of the YSR energies oscillates with a different amplitude as they originate from impurities sitting at different distances from the boundary.
The various parameters are chosen as $\tilde{J}=1.0 t$, $N=1000$, $\mu_l=-1.9t$, and $\Delta=0.005t$. }
\label{imp2}
\end{figure}

Next, we investigate how the closeness to the boundary affects the YSR energies of two magnetic impurities and, thus, the magnetic ground state of the system. In previous studies, it has been observed that, far away from the boundary, the total energy of such a system is extremized when the impurities are collinear~\cite{17}. In this study, we also focus on ferromagnetic and antiferromagnetic configurations. 
First,  we consider two impurities aligned ferromagnetically with equal exchange coupling strengths.
Due to the spatial overlap between the YSR states created by the two impurities, their energy levels split, lifting the initial twofold degeneracy \cite{17,23}. We numerically calculate the energy of the hybridized YSR states for different positions of the impurities with respect to the boundary, keeping the inter-impurity distance fixed to the value $x_{12}=28a$ for  $\tilde{J}=1.0 t$. In Fig.~\ref{imp2}(a), the energy $E_s$ of two YSR states is plotted as a function of the distance  $x_1/a$. We find that the boundary effects influence the hybridization  between the YSR states leading to oscillation of the energy levels with a $\lambda_F/2$ periodicity, similar to the single impurity system. As the impurity-boundary distance increases, the amplitude of oscillation decays exponentially with a decay length of $\xi_{sc}$. In the case of AFM orientation, the YSR  wavefunctions are orthogonal to each other, which keeps the YSR energy levels degenerate when impurities are far from the boundary. As the impurities approach the boundary, the YSR energies corresponding to each impurity exhibit oscillations with a periodicity of $\lambda_F/2$ caused by the boundary effects described above in the case of a single impurity. The amplitudes of the oscillations of the YSR energies are different as the impurities are located at different distances from the boundary. The presence of boundary, thus, lifts the degeneracy of the YSR energies as shown in Fig.~\ref{imp2}(b). 

\subsection{Magnetic Ground State of Two Impurities}
\label{twoimpgr}

\begin{figure*}[!t]
\begin{center}
\hspace{-.4cm}
\subfigure[]{\ig[width=4.9cm]{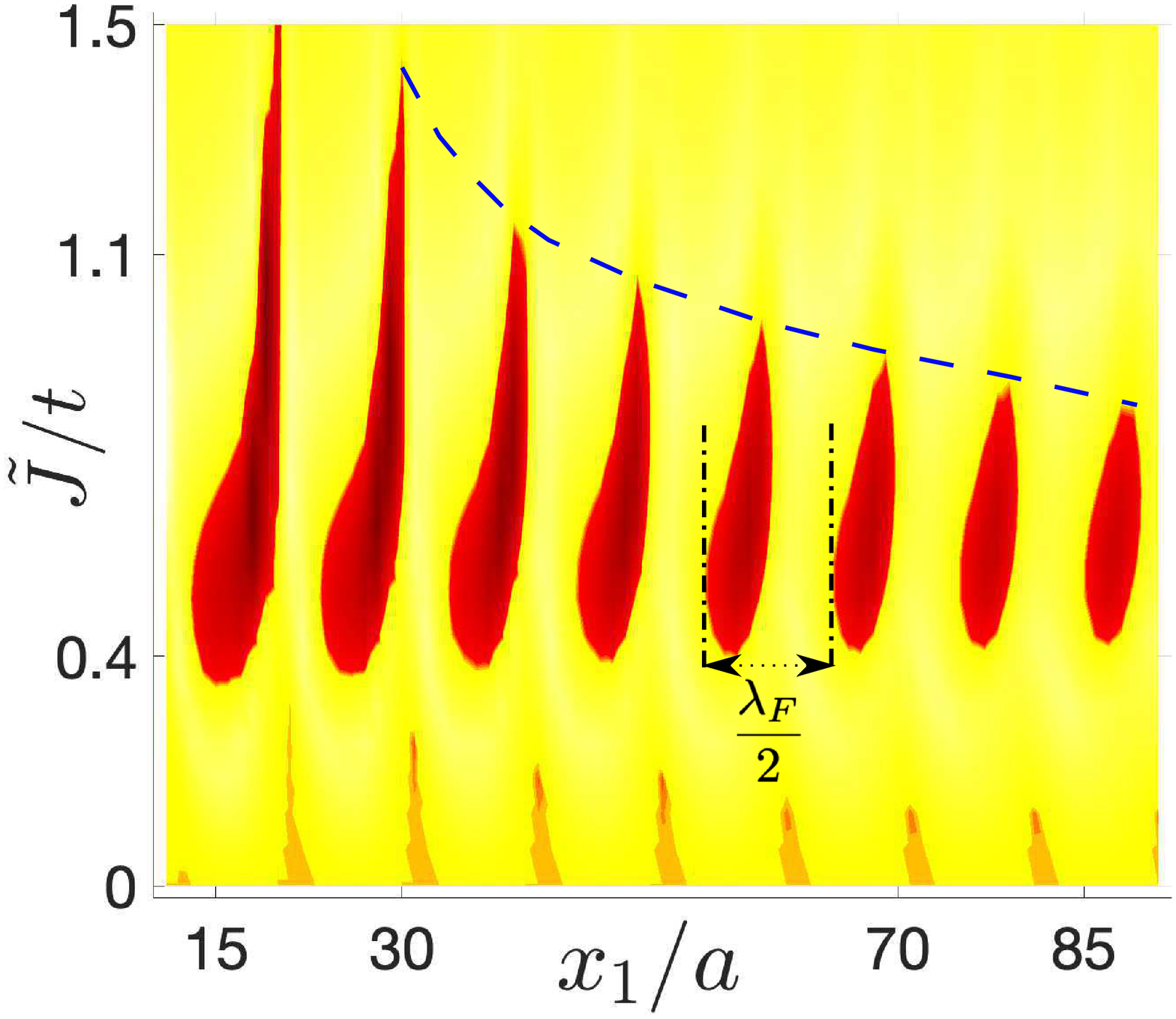}} 
\subfigure[]{\ig[width=5.2cm]{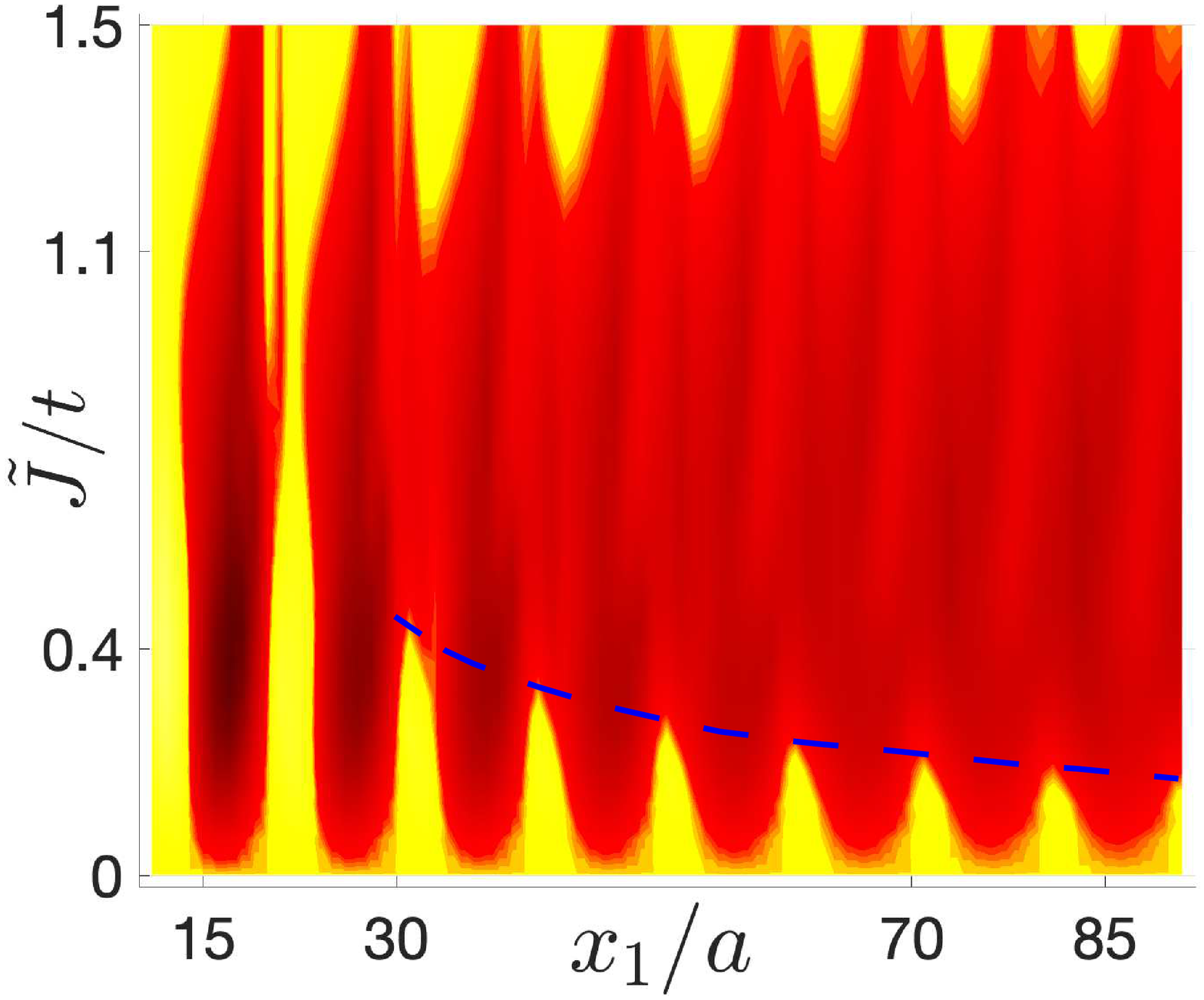}}
\subfigure[]{\ig[width=5.2cm]{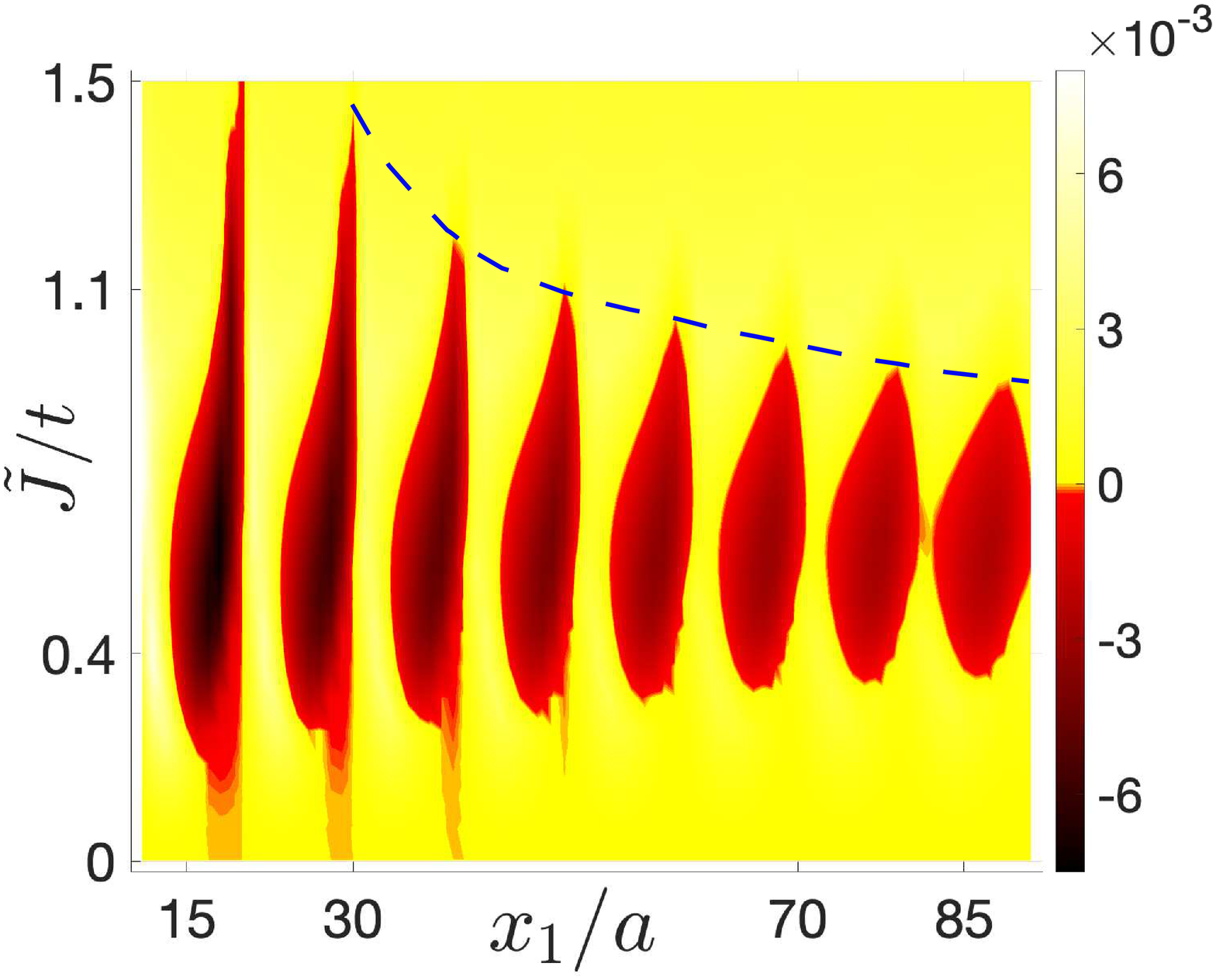}} \\ 
\hspace{-.45cm}
\subfigure[]{\ig[width=5.1cm]{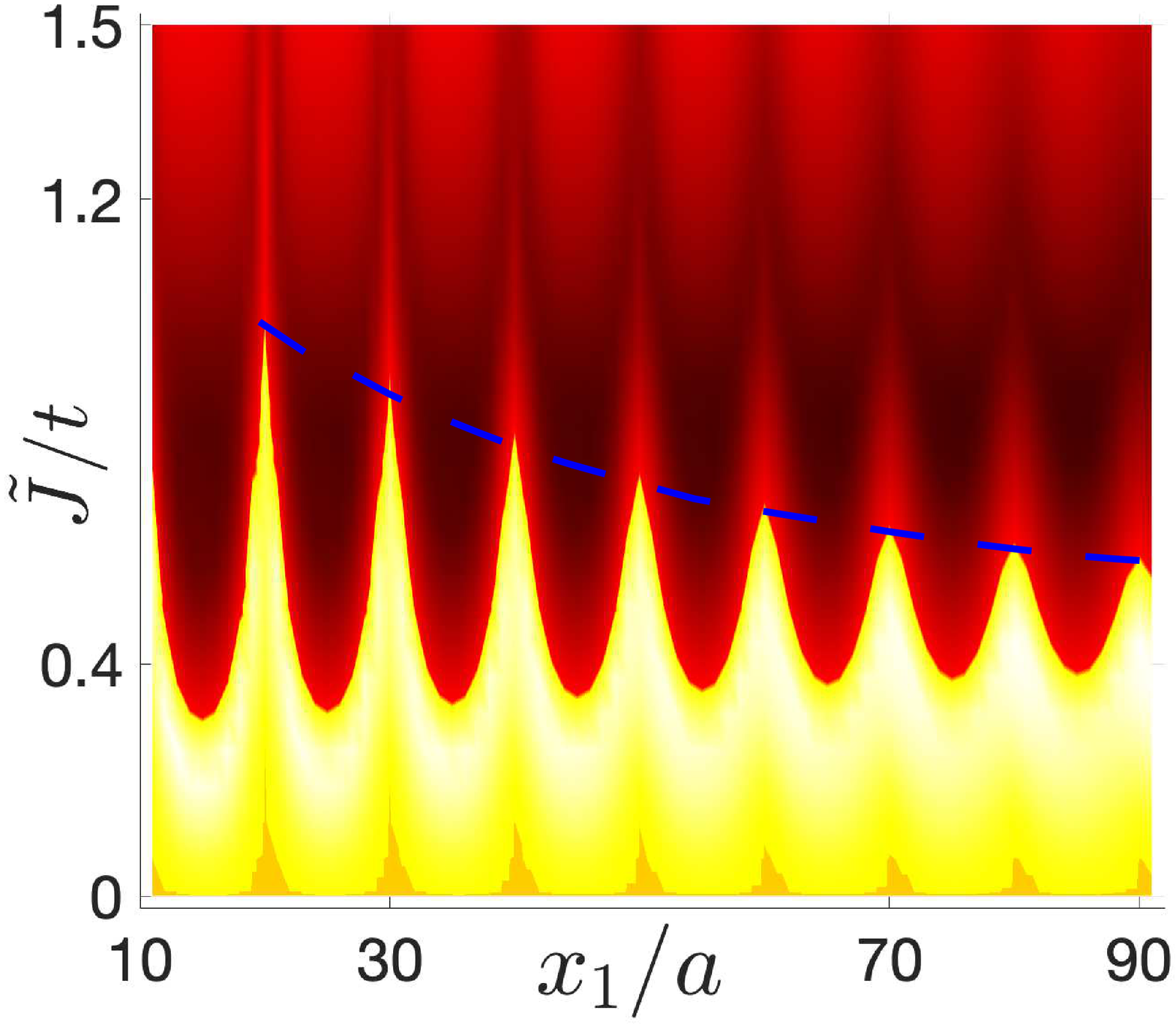}} 
\hspace{.05cm}\subfigure[]{\ig[width=5.15cm]{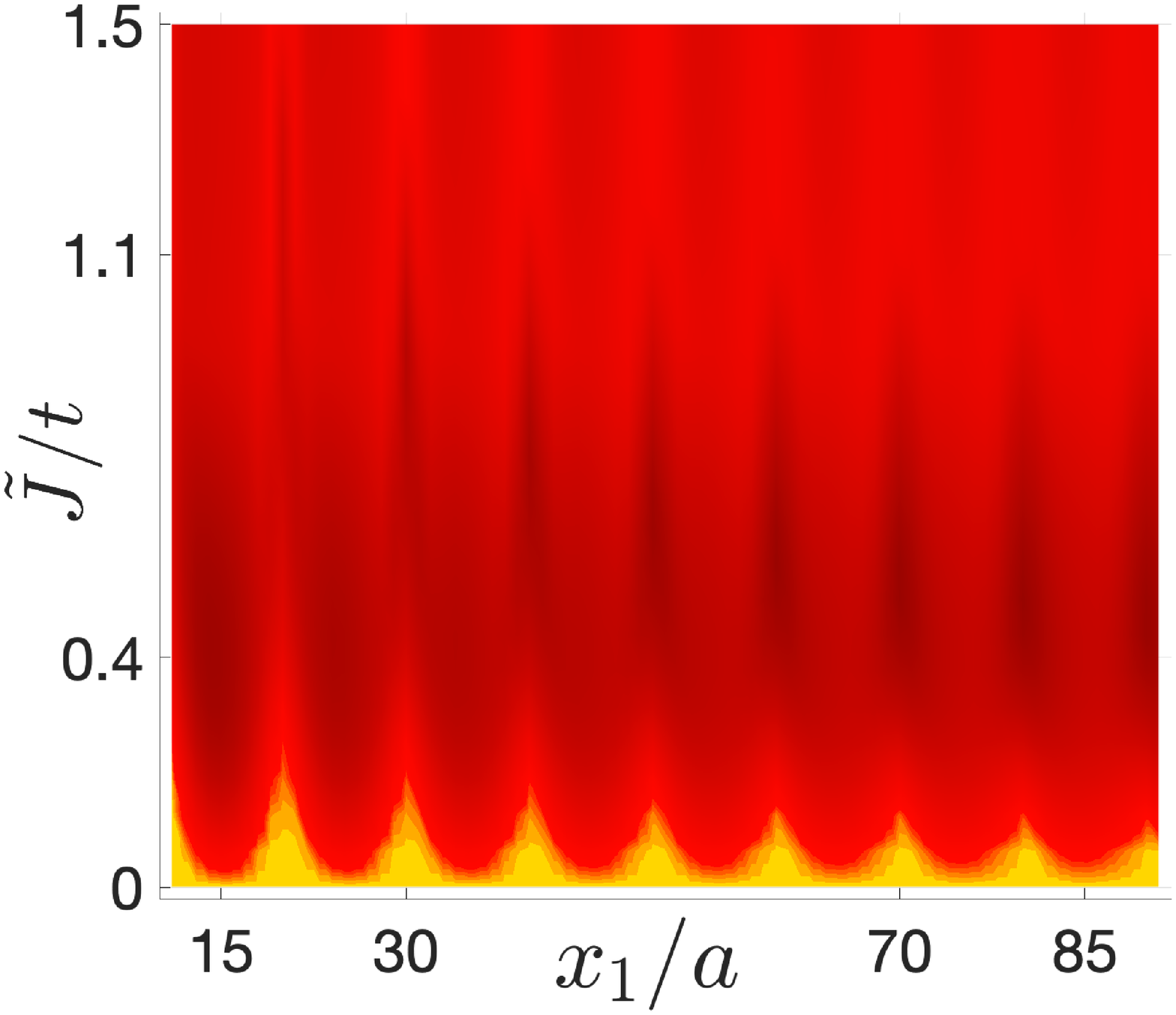}} 
\hspace{.05cm}\subfigure[]{\ig[width=5.23cm,keepaspectratio]{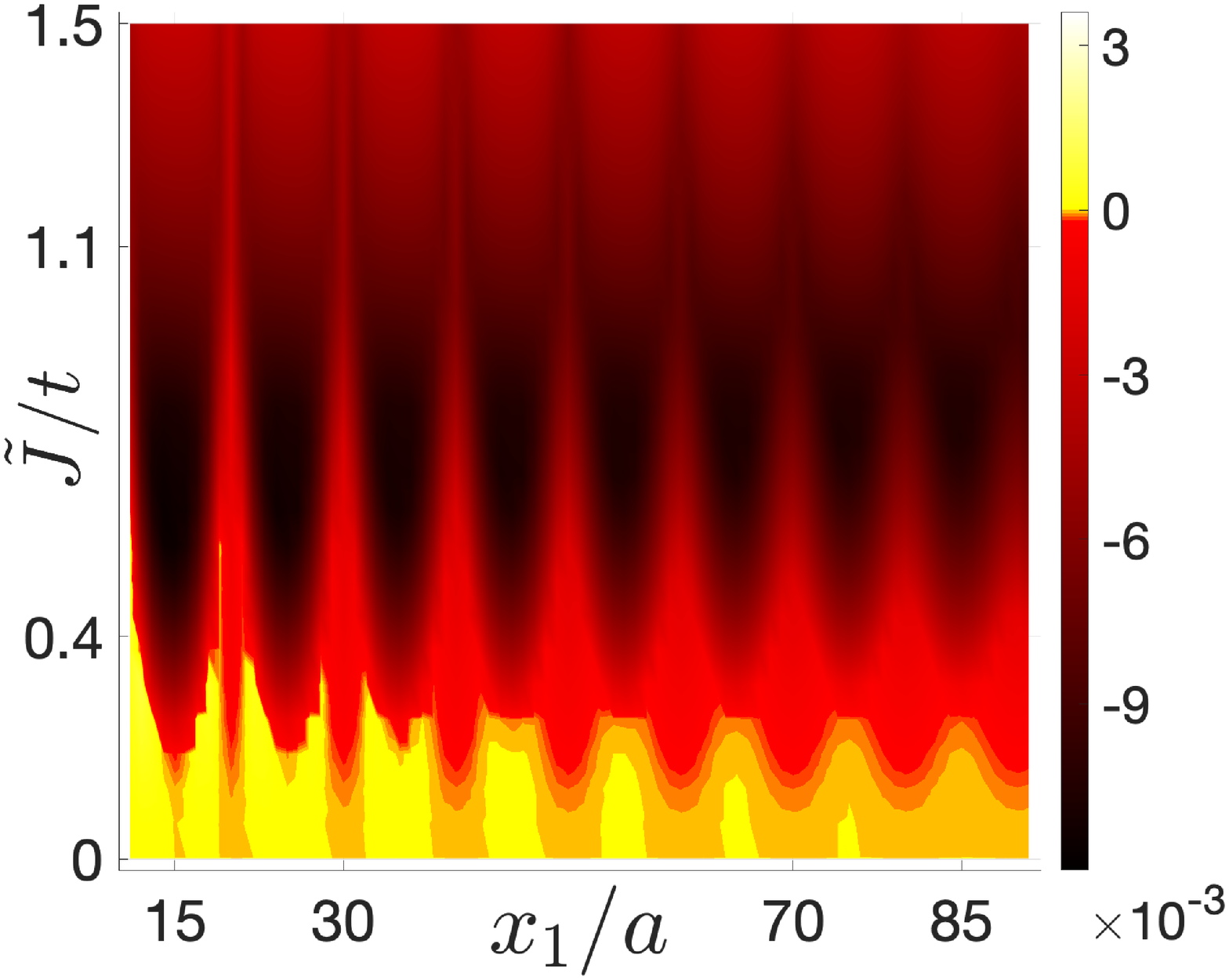}} 
\caption{The energy differences $\delta E_{\textrm{YSR}}$ (left panel), $\delta E_{\textrm{qp}}$ (middle panel), and $\delta E_{\textrm{g}}$ (right panel) between FM and AFM configuration as a function of $x_1/a$ along the $x$ axis and $\tilde{J}/t$ along the $y$ axis. In the first and second row of the figure, the inter-impurity distance is $x_{12}=28 a$ and $40 a$, respectively, such that $2k_F x_{12} $ is approximately an odd integer multiple of $\pi/2$ in the first row and an integer multiple of $\pi$ in the second row.  The red (yellow) patches indicate a FM (AFM) configuration of impurities. The magnetic configuration alternates between the FM and AFM phases with a periodicity of $\lambda_F/2$. The amplitude of oscillations decay exponentially as a function of $x_1$ due to the presence of the superconducting gap $\Delta$. The blue dashed line in the figures denote the exponentially decaying envelope function. The phase diagram for the total magnetic ground state is a sum of the contributions coming from the YSR and the quasiparticle states. Evidently, the oscillations in $\delta E_{\textrm{g}}$ is observed for a larger range of $\tilde{J}$ in Fig. \ref{fig:ediff}(c) than in Fig. \ref{fig:ediff}(f) in the second row, indicating the stronger effect of boundary when $2k_F x_{12}$ is an odd-integer multiple of $\pi/2$. The various parameters are chosen as $\mu=-1.9t$, $\Delta=0.005t$, and $N=500$.}
\label{fig:ediff}
\end{center}
\end{figure*}

The ground state energy of the system is dependent on the relative angle $\theta$ between the magnetic impurities. The total energy of the system
 $E_{\textrm{g}}(\theta)=E_{\textrm{qp}}(\theta)+E_{\textrm{YSR}}(\theta)$ is calculated by summing over all the negative energy states, i.e. all the energies below the chemical potential \cite{9,15}. Generally, $E_{\textrm{g}}(\theta)$ can be divided into two contributions: $E_{\textrm{YSR}}(\theta)$ coming from the YSR states and $E_{\textrm{qp}}(\theta)$ coming from the quasiparticle states. The ground state energy difference between the collinear configurations of the impurities is given by $\delta E_{\textrm{g}} = E_{\textrm{g}}(0) - E_{\textrm{g}}(\pi)$ where $\theta=0$ ($\theta= \pi$) denotes a FM (AFM) configuration, respectively. For $\delta E_{\textrm{g}} > 0$ ($\delta E_{\textrm{g}} < 0$), the ground state of the system is AFM (FM). In this section, we investigate how the ground state configuration depends on the impurity positions $x_1$ and $x_2$ as well as on the exchange interaction strength $\tilde J$.  We also pay  attention to the relative contributions of the YSR and quasiparticle states. The former should dominate at longer distances since the YSR states do not have a power-law decay in one dimension. As such, we analogously define $\delta E_{\textrm{YSR}}=E_\textrm{YSR}(0) - E_\textrm{YSR}(\pi)$ and $\delta E_{\textrm{qp}}=E_\textrm{qp}(0) - E_\textrm{qp}(\pi)$.

In Fig.~\ref{fig:ediff}, we plot $\delta E_{\textrm{YSR}}$ [panels (a) and (d)], $\delta E_{\textrm{qp}}$ [panels (b) and (e)], and $\delta E_{\textrm{gr}}$ [panels (c) and (f)] as a function of $x_1$ and $\tilde J$, for two different values of the inter-impurity distance $x_{12}$; the preference of the FM (AFM) ground state is indicated by red (yellow) patches in the figure. In the first row of Fig.~\ref{fig:ediff}, we choose $x_{12}=28 a$ such that $2k_F  x_{12} \approx (2\times n+1)\pi/2$ with $n=5$.  According to our analysis of the RKKY interaction in Sec.~\ref{twoimp}, the boundary effects should be relatively strong for $x_{12}$ satisfying such a condition.  Here, we investigate numerically  such boundary effects for various values of the exchange interaction strength $\tilde{J}$. In Fig.~\ref{fig:ediff}(a), the magnitude of $\delta E_{\textrm{YSR}}$ is negligibly small for small $\tilde{J}$. With increasing $\tilde{J}$, we observe oscillations in $\delta E_{\textrm{YSR}}$ between the FM (red) and the AFM (yellow) configurations as a function of $x_1$ and with a periodicity of $\lambda_F/2$. Further increase in $\tilde{J}$ reduces such oscillations in $\delta E_{\textrm{YSR}}$ as a function of $x_1$. Moreover, the relative size of the red (FM) region shrinks as the impurities move away from the system boundary as shown in Fig.~\ref{fig:ediff}.

The salient features in the Fig.~\ref{fig:ediff}(a) can be understood as follows by considering the energies of the YSR states: When $\tilde{J}$ is small, the YSR energies lie close to the gap edge, and, consequently, the energy difference between the FM and AFM configurations is small. With increasing $\tilde{J}$ the YSR states move deeper inside the superconducting gap. The boundary-induced hybridization between these states then results into oscillations in the YSR energies around zero as a function of $x_1$, thereby giving rise to phase transitions between the FM and the AFM configurations with varying impurity-boundary distances \cite{17}. As we keep on increasing $\tilde{J}$, the YSR energies move back towards the gap edge, thus again reducing the oscillations between the FM and the AFM phases. To analyze the decay in the relative phase space of the FM phases as $x_1$ increases, we perform a curve fitting of the envelope function [shown by the blue dashed curve in Fig.~\ref{fig:ediff}(a)] obtained by connecting the topmost points of the phase boundary between the red and yellow regions. We find that the envelope is an exponentially decaying function of $x_1$ with a decay length of order of $\xi_{sc}$. This feature originates from the exponential decay of the boundary-induced hybridization between the YSR states as discussed in Sec.~\ref{hyb}.

Next, in Fig.~\ref{fig:ediff}(b), we plot $\delta E_{\textrm{qp}}$ to analyze the contribution of the quasiparticle states to the total energy. Here too, the magnetic configuration oscillates between the FM (red) and AFM (yellow) phases as a function of $x_1$. For small values of $\tilde{J}$, we observe strong boundary-induced oscillations in $\delta E_{\textrm{qp}}$ around zero. As we increase $\tilde{J}$, the oscillations between FM and AFM phases get suppressed and the magnetic configuration becomes almost independent of the impurity-boundary distance $x_1$. With further increase in $\tilde{J}$, $\delta E_{\textrm{qp}}$ again exhibits oscillation around zero as a function of $x_1$, similar to that in the small $\tilde{J}$ limit. 

To understand the origin of the oscillations in $\delta E_{\textrm{qp}}$, we first note that at small values of $\tilde{J}$, the exchange interaction between the impurities is of the RKKY type. At inter-impurity distances satisfying the condition $2 k_F x_{12}=(2n+1)\pi/2$, the RKKY Hamiltonian in Eq. \eqref{eq:hbrkky} is independent of $x_{12}$ and depends only on the boundary induced RKKY coefficients as described in Sec.~\ref{twoimp}, thus giving rise to the strongest boundary effect for this choice of $x_{12}$. The phase oscillations have a periodicity of $\lambda_F/2$ as also predicted by the RKKY interaction. The envelope function marked by the blue dashed line in Fig.~\ref{fig:ediff}(b) exhibits exponential decay as a function of $x_1$, induced by the presence of the  superconductivity. We also obtained that for very large values of $\tilde{J}$, the oscillation amplitude of $\delta E_{\textrm{qp}}$ decays and the preferable magnetic configuration is an AFM ordering, independent of $x_1$ values.

Finally, in Fig.~\ref{fig:ediff}(c) we plot the energy difference $\delta E_{\textrm{g}}$ including both the subgap YSR states and the bulk quasiparticle states. The phase diagram is a composite of the contributions coming from both YSR states and quasiparticles. Evidently, the phase boundary between the magnetic configurations is largely determined by the YSR states \cite{14}. This arises from the fact that at $2 k_F x_{12}=(2n+1)\pi/2$, the strongest term $F(x_{12})$ in the RKKY interaction expression, mediated by the quasiparticle states, goes to zero. As a result, the boundary-induced effect in the RKKY interaction is substantially overpowered by the contribution from the YSR states, which do not have any power law decay prefactor in one-dimensional systems. In Fig.~\ref{fig:ediff}(c), for small values of $\tilde{J}$, the oscillation in $\delta E_{\textrm{g}}$ around zero is negligible, showing a weak effect of the boundary on the magnetic ground state. With increasing $\tilde{J}$, the ground state configuration begins to alternate with $\lambda_F/2$ periodicity, as also seen in the phase diagram of $\delta E_{\textrm{YSR}}$ in Fig.~\ref{fig:ediff}(a).  We observe an expansion of the FM ground state in phase space compared to that in Fig.~\ref{fig:ediff}(a), arising from the bulk contributions to the total magnetic ground state.  As $\tilde{J}$ is increased further, the oscillations in $\delta E_{\textrm{g}}$ decrease similar to the small exchange limit and an AFM orientation of the impurities is favoured. We find that for large $x_1$, the phase diagram is mostly determined by the YSR states since the quasiparticle contributions decay as a power law with increasing impurity-boundary distances whereas the YSR contribution does not. Here also, the envelope function denoted by the blue dashed curve in the figure is an exponentially decaying function of $x_1$. Our results re-emphasize the fact that the YSR states play a crucial role in determining the correct ground state configuration, thereby making it important to include both the YSR states and the quasiparticle states while finding the magnetic ground state.

In the second row of Fig.~\ref{fig:ediff}, we choose $x_{12}=40a$ such that $2k_F x_{12}=n\times\pi$ with $n=2$. In this regime of $x_{12}$, the effect of boundary on the ground state configuration is minimal in the RKKY limit discussed in Sec. \ref{twoimp}. In Fig.~\ref{fig:ediff}(d) for $\delta E_{\textrm{YSR}}$, the energy difference between the FM and the AFM configurations is negligible at small values of $\tilde{J}$, similar to the phase diagram in Fig.~\ref{fig:ediff}(a) for $x_{12}=28a$. With increasing $\tilde{J}$, the energy difference exhibits oscillations around zero with a significant amplitude and a periodicity of $\lambda_F/2$. These oscillations in $\delta E_{\textrm{YSR}}$ arise from the boundary-induced hybridization of the YSR states whose energies lie deep inside the superconducting gap at such $\tilde{J}$ values. With further increase in $\tilde{J}$, the oscillations in the magnetic configuration become negligible as the YSR states go back to the gap edge, similar to the YSR physics discussed for $x_{12}=28 a$. The envelope (blue dashed line) obtained by connecting the critical points of the phase boundary as discussed above is an exponentially decaying function of $x_1$, as also observed in the previous regime.

In Fig.~\ref{fig:ediff}(e), we show the energy contribution $\delta E_{\textrm{qp}}$ coming from the quasiparticle states. We do not observe any oscillation in $\delta E_{\textrm{qp}}$ around zero as a function of $x_1$, indicating a suppression of the boundary effect. To analyze this, we first recall that for small values of $\tilde{J}$, the exchange interaction between the impurities is governed by the RKKY Hamiltonian in Eq.~\eqref{eq:hbrkky}. At inter-impurity distance satisfying $2k_F x_{12}=n\pi$, the $x_{12}$-dependent RKKY coefficient $F(x_{12})$ dominates over the boundary-induced coefficients $F(X)$ and $F(x_2)$ in Eq.~\eqref{eq:hbrkky}, thereby suppressing the boundary effect at this regime of $x_{12}$. We checked our results for very large values of $\tilde{J}$ and find that such behavior of $\delta E_{\textrm{qp}}$ is not limited to small $\tilde{J}$ values but instead holds for the entire range of $\tilde{J}$ within the checked parameter range.

Finally, in Fig.~\ref{fig:ediff}(f), we plot the energy difference $\delta E_{\textrm{g}}$ calculated from the total energies of two competing magnetic ground states. In this regime too, both the YSR states and the quasiparticle states contribute significantly to the total energy $\delta E_{\textrm{g}}$. For small $\tilde{J}$ values, $\delta E_{\textrm{g}}$ exhibits some oscillations between the FM and the AFM ground states as a function of $x_1$ but these oscillations vanish completely upon increasing $\tilde{J}$. The oscillations in $\delta E_{\textrm{g}}$ originate from the interplay between the YSR and the bulk contributions to the total magnetic ground state. Unlike the previous regime where the YSR contribution tends to dominate over the bulk, here the bulk contribution is of the same order as the YSR contribution since the strongest term $F(x_{12})$ in the RKKY interaction attains the maximum magnitude at this regime. Therefore, the phase boundary in the phase diagram of $\delta E_{\textrm{g}}$ does not closely follow that of $\delta E_{\textrm{YSR}}$. As $\tilde{J}$ increases, the bulk contribution overpowers the YSR contribution, leading to the decay in the oscillations in $\delta E_{\textrm{g}}$. We also observe that the oscillations in the ground state configuration at small values of $\tilde{J}$ vanish with increasing $x_1$. This arises from the fact that the boundary dependent RKKY coefficients $F(X)$ and $F(x_2)$ decay with increasing $x_1$, whereas $F(x_{12})$ remains unaltered, resulting in an increasing magnitude of $F_{\textrm{sum}}=F(x_{12})+F(X)-2F(x_{2})$ as a function of $x_1$. Hence, the bulk contribution begins to dominate over the YSR contribution leading to the decay in the oscillations in $\delta E_{\textrm{g}}$ as the impurities are moved away from the boundary.

\begin{figure*}[!tb]
\begin{center}
\subfigure[]{\ig[width=5.5cm]{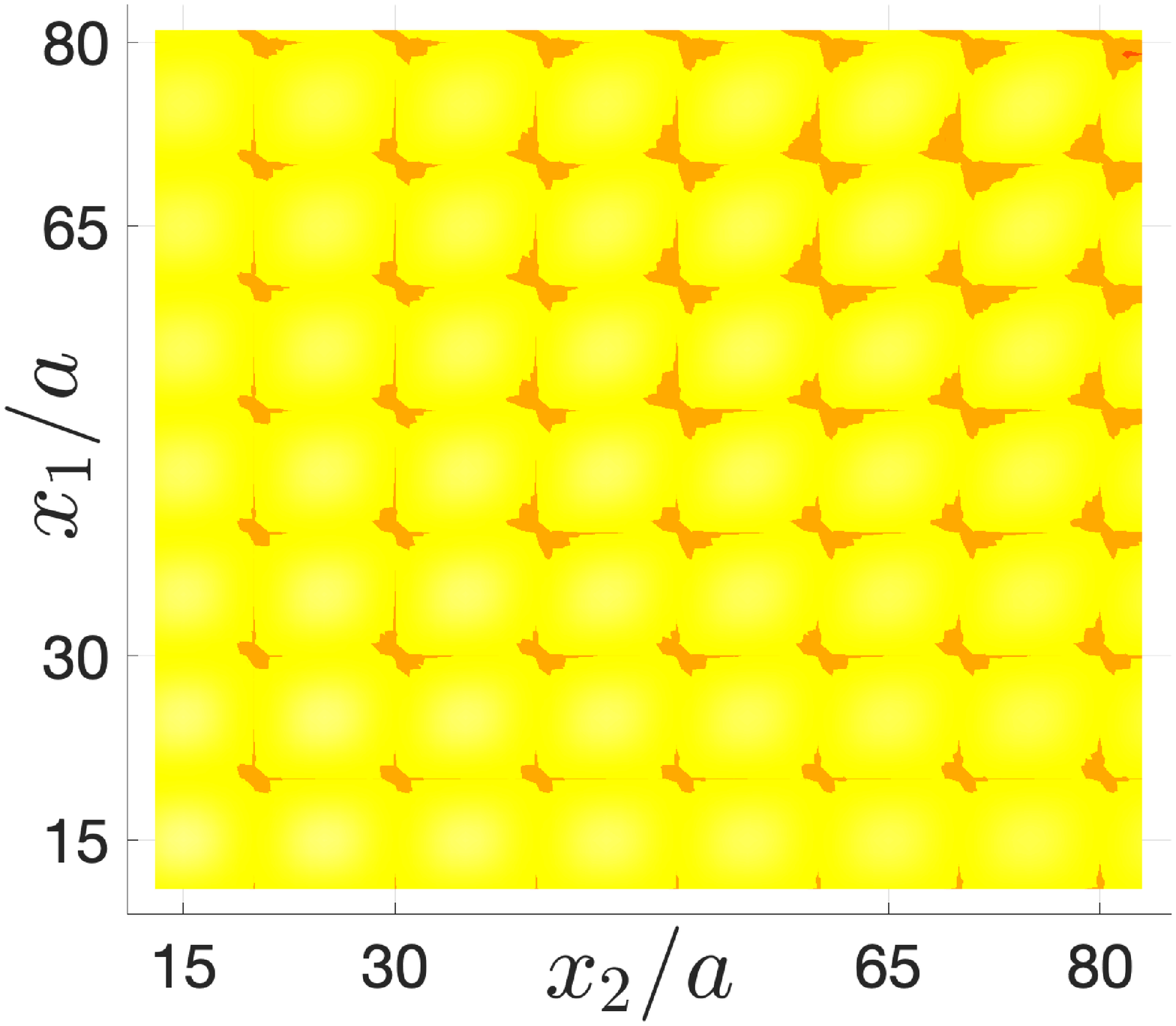}} 
\subfigure[]{\ig[width=5.5cm]{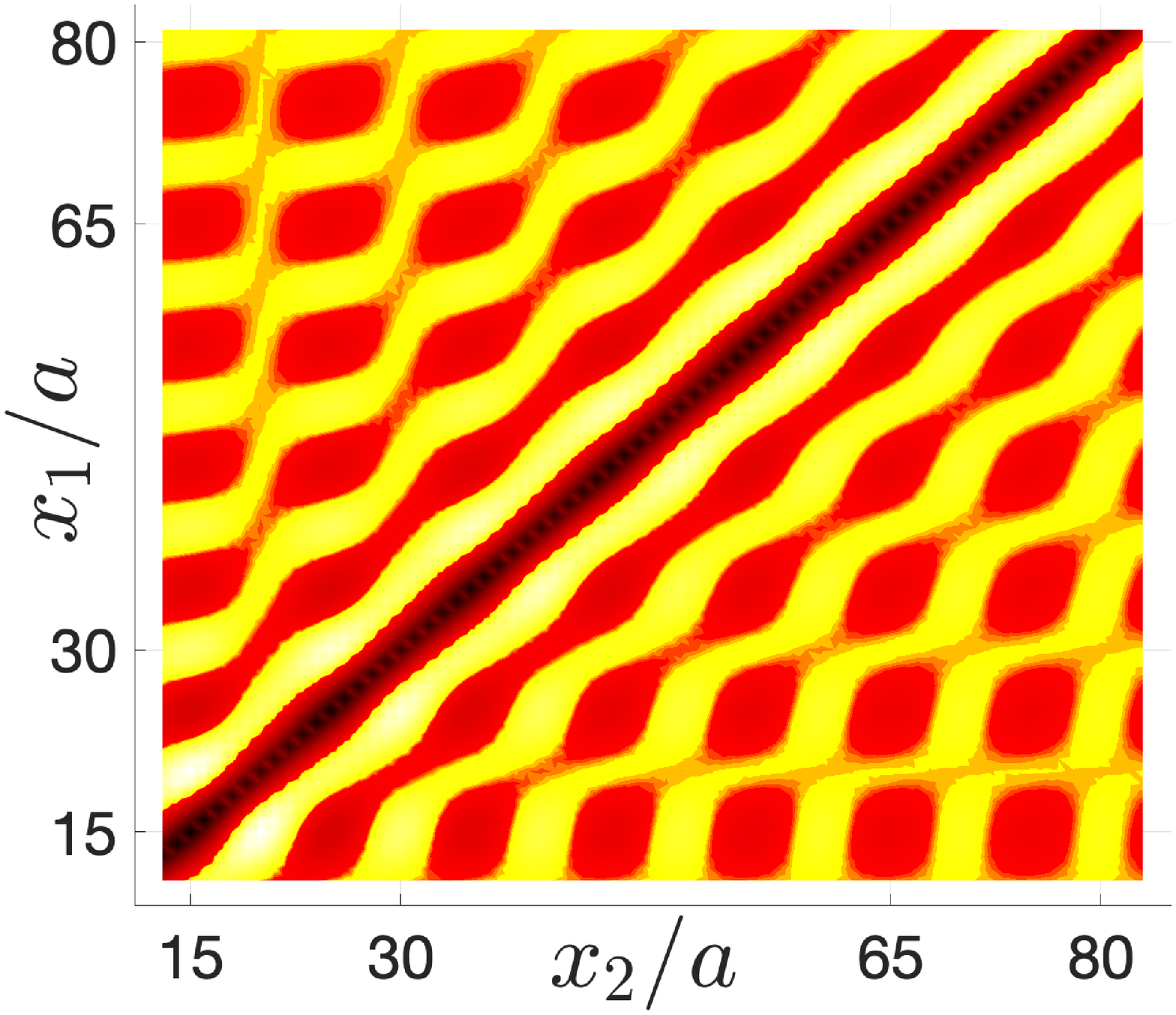}}
\subfigure[]{\ig[width=5.6cm]{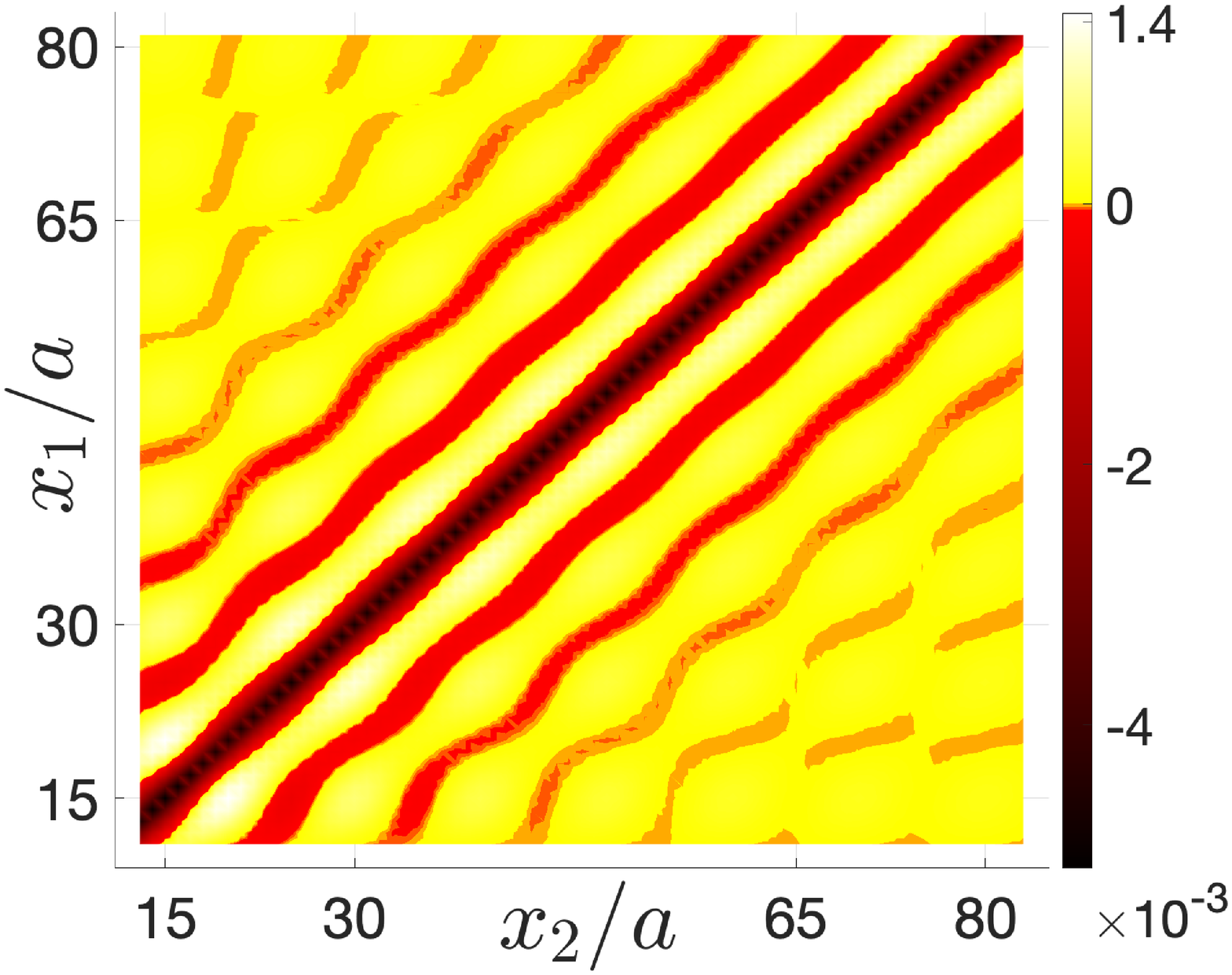}} \\ 
\subfigure[]{\ig[width=5.5cm]{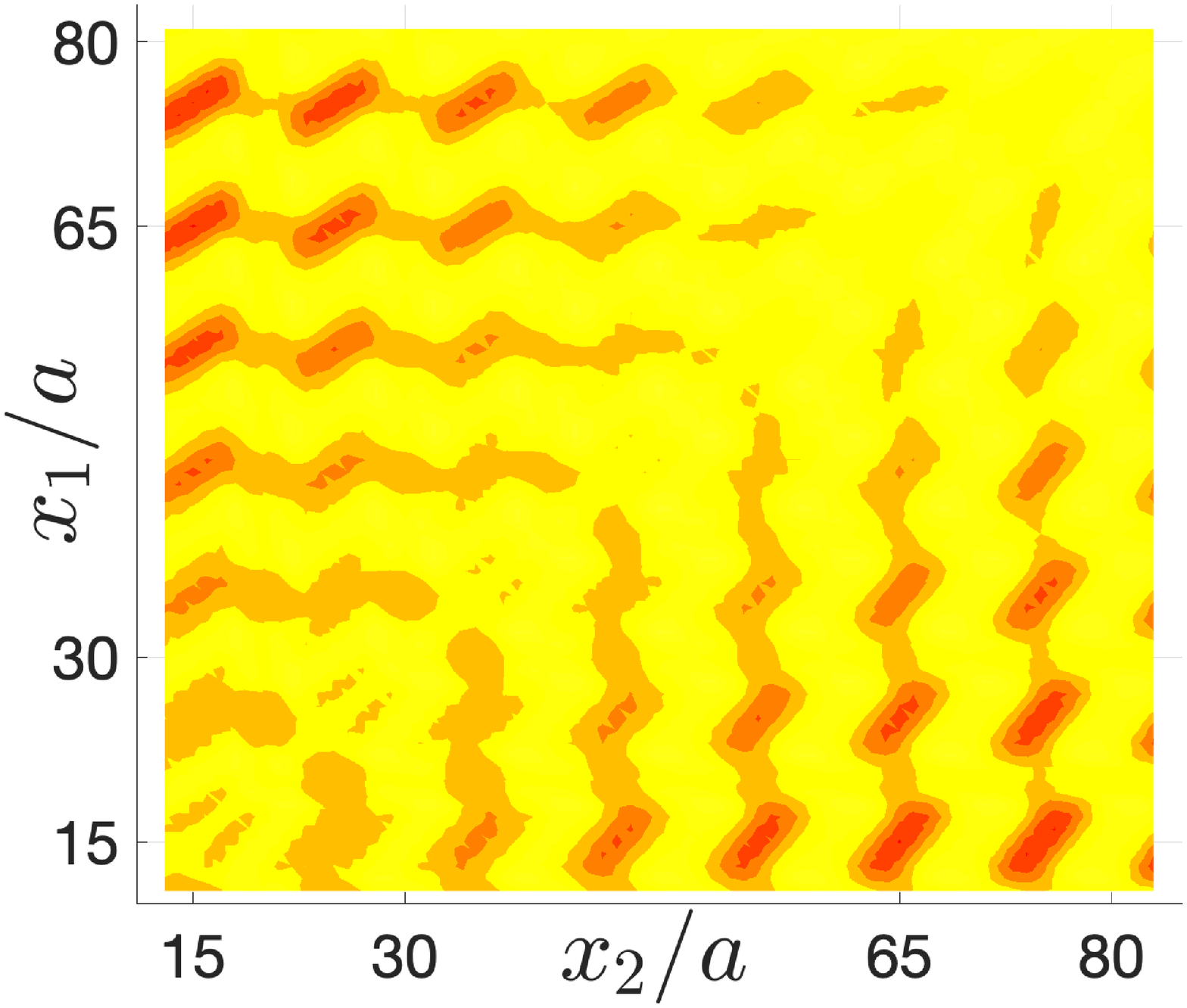}} 
\hspace{.05cm}\subfigure[]{\ig[width=5.5cm]{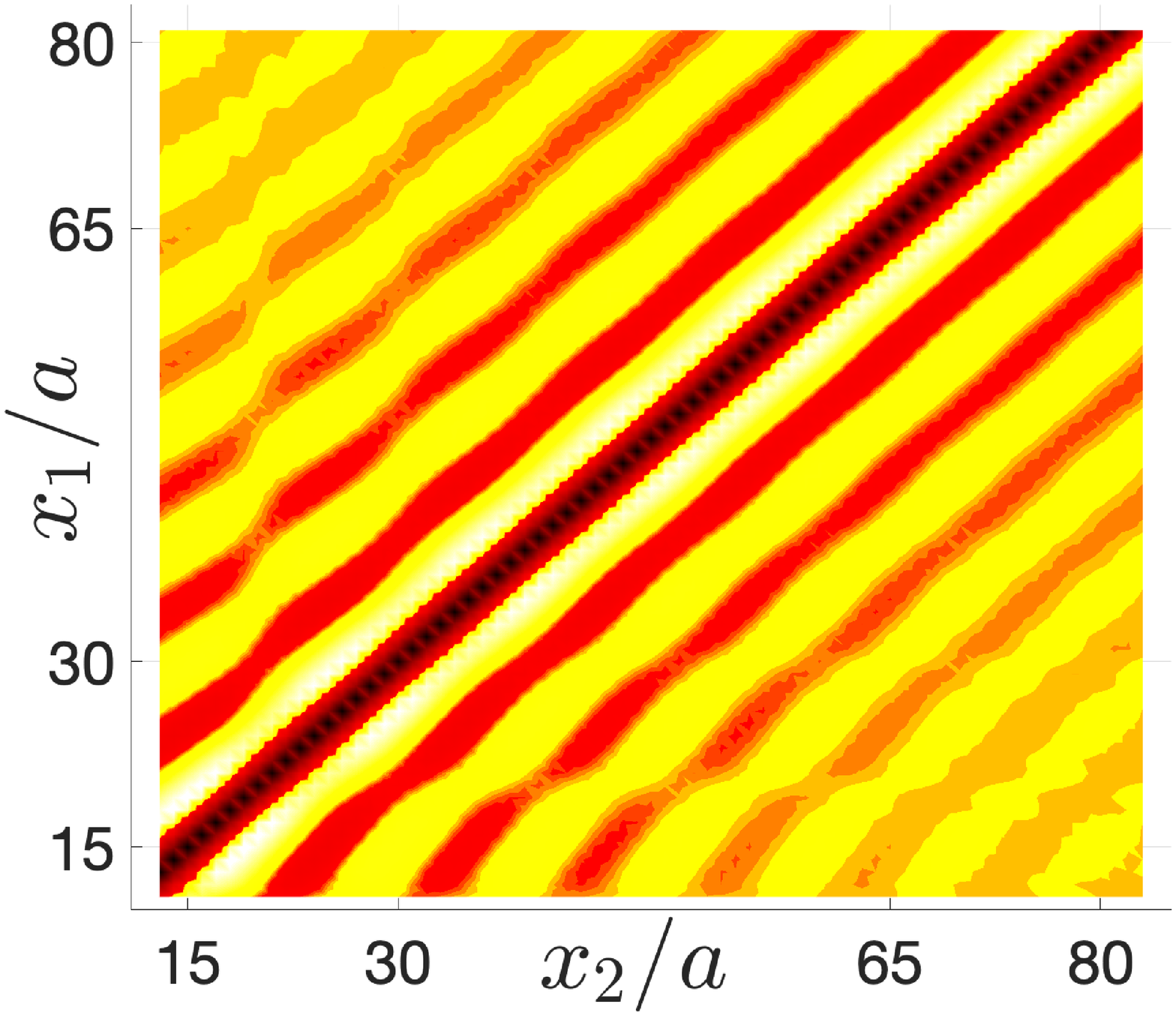}} 
\subfigure[]{\ig[width=5.6cm]{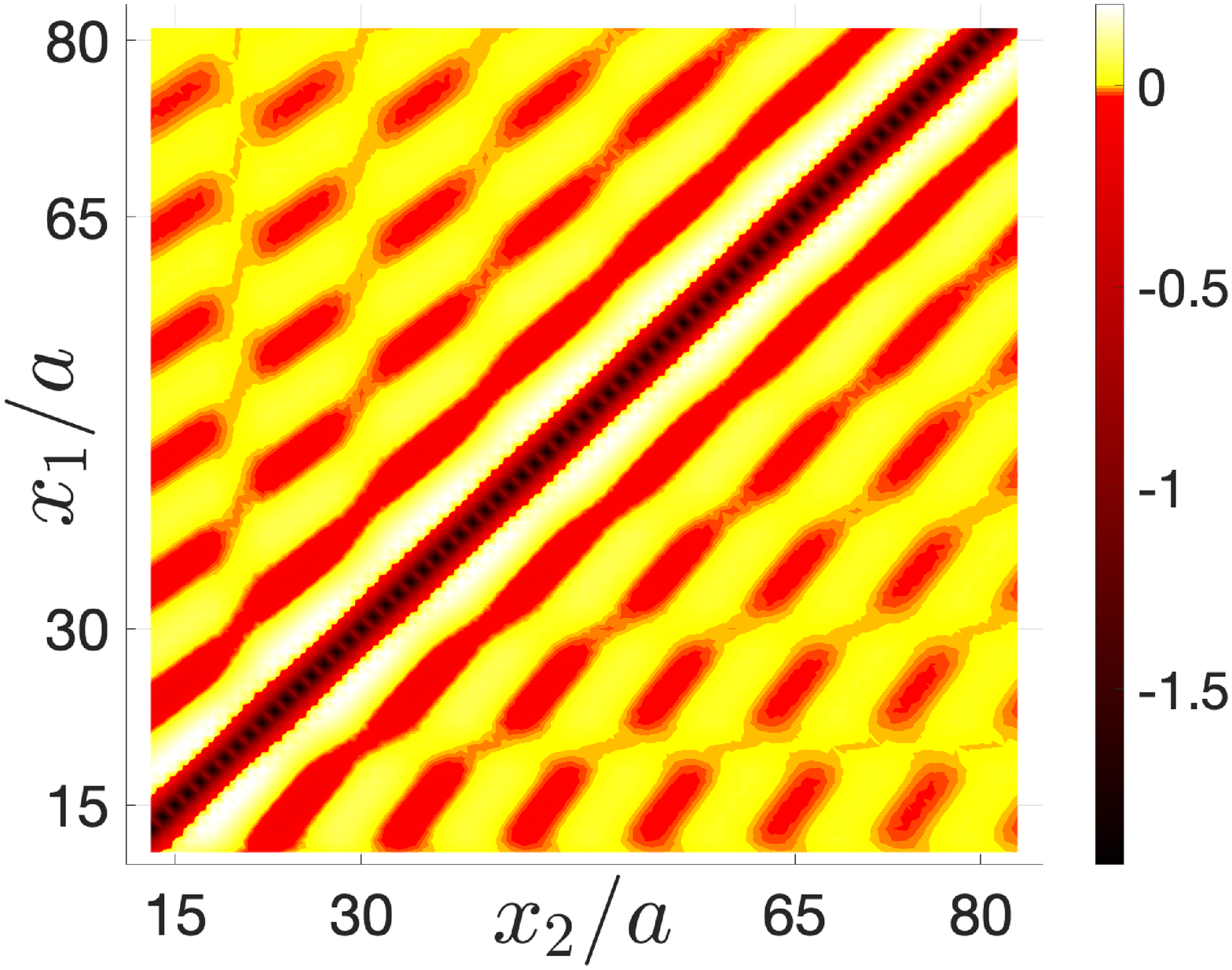}} \\
\subfigure[]{\ig[width=5.5cm]{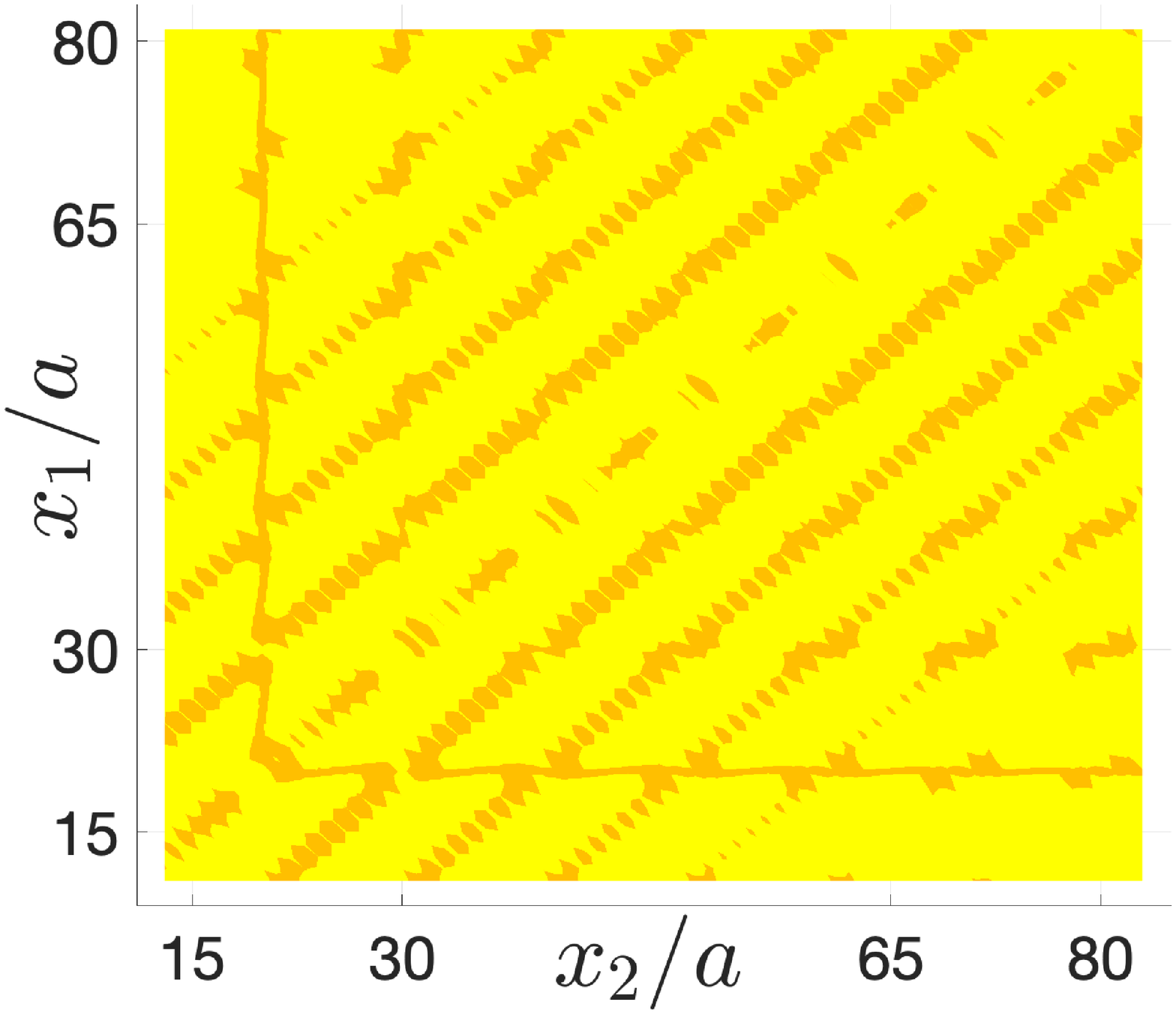}} 
\hspace{.05cm}\subfigure[]{\ig[width=5.5cm]{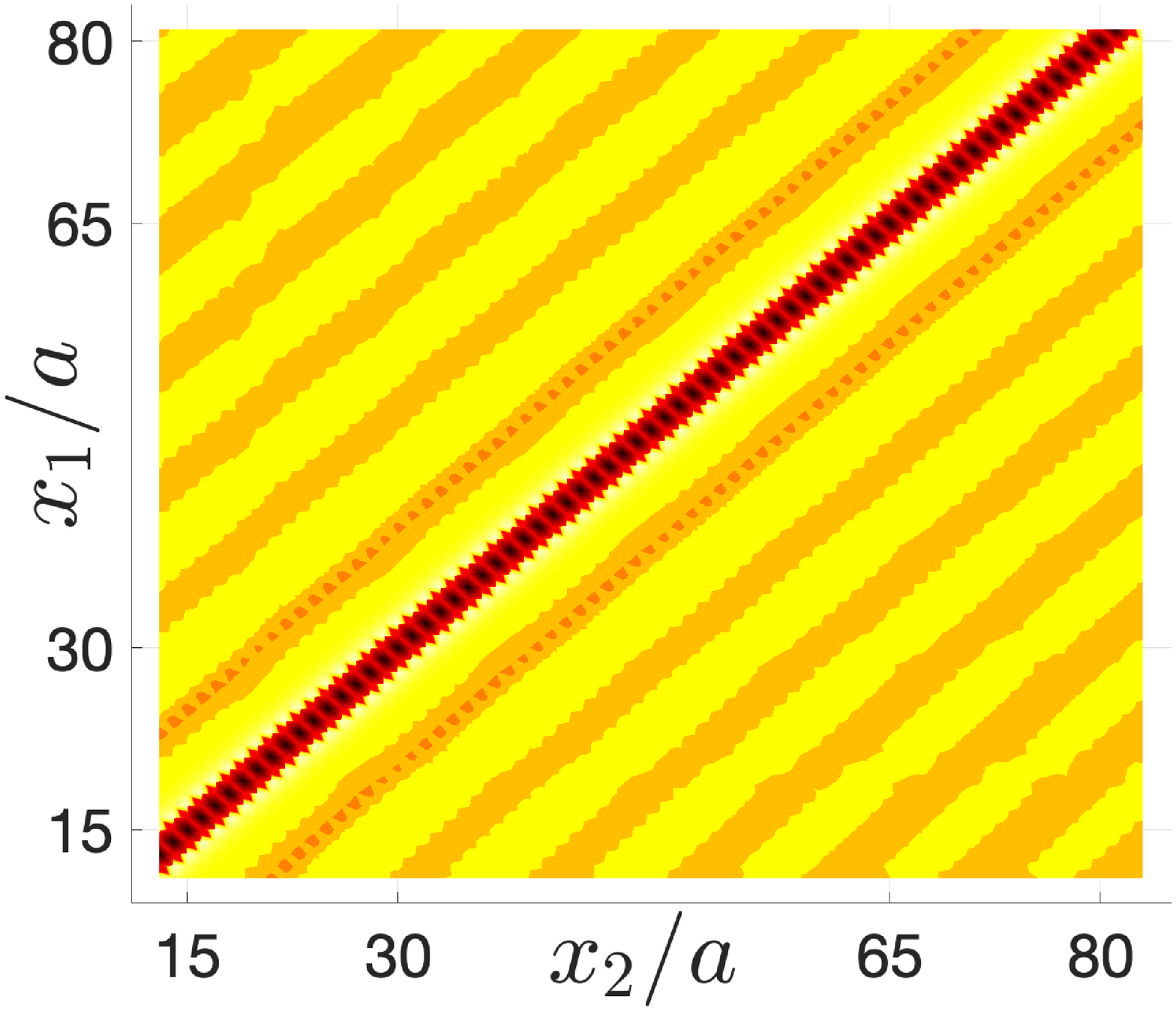}} 
\subfigure[]{\ig[width=5.6cm]{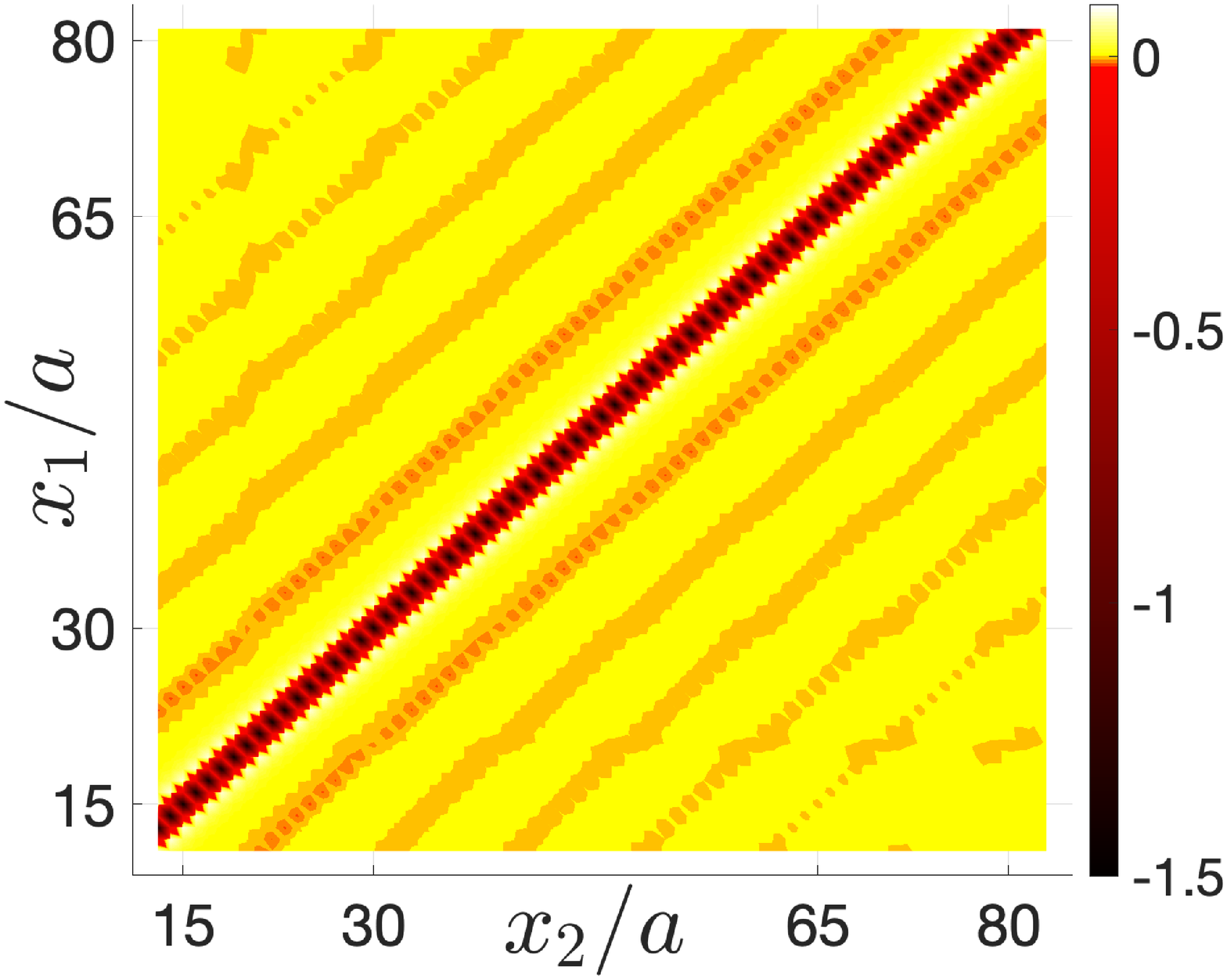}} 
\caption{The energy differences $\delta E_{\textrm{YSR}}$ (left panel), $\delta E_{\textrm{qp}}$ (middle panel) and $\delta E_{\textrm{g}}$ (right panel) as a function of $x_1/a$ along the $x$-axis and $x_2/a$ along the $y$-axis. In the first, second, and third row, $\tilde{J}=0.05t$, $0.5t$, and $1.5t$ respectively. The red regions in the phase diagrams indicate a FM configuration  while the yellow regions indicate an AFM one. The energy difference $\delta E_{\textrm{YSR}}$ exhibits oscillations around zero along the lines of constant $x_{12}$, originating from the boundary-induced hybridization between the YSR states. The plots for $\delta E_{\textrm{qp}}$ also show similar oscillations except close to the region $x_{12}\approx 0$.  Both the YSR states and the quasiparticle states contribute significantly in the phase diagram of $\delta E_{\textrm{g}}$ for the total magnetic ground state. When $|x_{12}|<\lambda_F$, the phase diagram of $\delta E_{\textrm{g}}$ is governed by the quasiparticle contribution, while beyond this regime, the YSR  contribution dominates.  The other parameters are fixed as $\mu=-1.9t$, $\Delta=0.005t$, and $N=500$. }
\label{fig:ediff2}
\end{center}
\end{figure*}

Next, to explore how the ground state configuration deviates from the one predicted by the RKKY analysis (see Fig.~\ref{fig:rkky}), we calculate the magnetic ground state as a function of $x_1$ and $x_2$ for $\tilde{J}=0.05t$ [weak coupling regime, Figs.~\ref{fig:ediff2}(a), (b), and (c)], $0.5t$ [intermediate coupling regime, Figs.~\ref{fig:ediff2}(d), (e), and (f)], and $1.5t$ [strong coupling regime, Figs.~\ref{fig:ediff2}(g), (h), and (i)]. The red and yellow patches denote the FM and AFM configurations, respectively. 

In the first row of the figure, we use $\tilde{J}=0.05t$. In Fig.~\ref{fig:ediff2}(a), the energy difference $\delta E_{\textrm{YSR}}$ for the YSR states is presented. Again, $\delta E_{\textrm{YSR}}$ exhibits oscillations around zero energy  along the lines of constant $x_{12}$, arising from the boundary-induced hybridization of the YSR states. The phase diagram has similar patterns along the constant $x_1$ and the constant $x_2$ lines, respectively, indicating the symmetric nature of the boundary effect as a function of the impurity-boundary distance. In Fig.~\ref{fig:ediff2}(b), we show the energy difference $\delta E_{\textrm{qp}}$ for the quasiparticle states at $\tilde{J}=0.05t$. We observe transitions between the FM and AFM configurations along the constant $x_{12}$ lines. However, close to the region $x_{12}\approx 0$, there is no transition in the magnetic configuration if $x_{12}$ is unaltered. These features look similar to that in the phase diagram in Fig.~\ref{fig:rkky} for the RKKY limit of the exchange interaction. This is due to the fact that at such a small value of $\tilde{J}$, the exchange interaction mediated by the quasiparticle states is of the RKKY type. 

Finally, in Fig.~\ref{fig:ediff2}(c), we plot the energy difference $\delta E_{\textrm{g}}$ between two magnetic ground states obtained by calculating the total energy of the system. The phase diagram is obtained from a sum of the YSR and the quasiparticle contributions, being mostly governed by the quasiparticle states for $|x_{12}|\lesssim \lambda_F$ and by the YSR states beyond that limit. As a result, we do not observe much oscillations in the ground state configuration along the constant $x_{12}$ lines around $x_{12}=0$, similar to that in Fig.~\ref{fig:ediff2}(b) for the quasiparticle states. With increasing inter-impurity distance, the RKKY interaction decays as $1/{k_F x_{12}}$ but the YSR states, not having any power law decay in 1D, contribute significantly in determining the total magnetic ground state. Thus, for $|x_{12}|>\lambda_F$, we observe oscillations in the ground state configuration along the constant $x_{12}$ lines, exhibiting a strong boundary effect in this regime. It is interesting to note that even for such a small $\tilde{J}$, the YSR states play an important role in determining the total magnetic ground state, thereby making it necessary to consider both the YSR and the quasiparticle states while calculating the magnetic ground state of impurities, even in the limit of weak exchange interaction.

In the second row of the figure, we choose $\tilde{J}=0.5t$. The energy difference $\delta E_{\textrm{YSR}}$, plotted in Fig.~\ref{fig:ediff2}(d), exhibits oscillations around zero along the constant $x_{12}$ lines except when $x_{12}\approx 0$. To understand this feature, first note that the YSR energies lie deep inside the superconducting gap at such a large value of $\tilde{J}$.  For inter-impurity distance $x_{12}\approx 0$, the strong hybridization between these YSR states push their energies close to the superconducting gap edge. Hence, the energy difference $\delta E_{\textrm{YSR}}$ does not exhibit any oscillation around zero along this constant $x_{12}$ line \cite{17}.  In Fig.~\ref{fig:ediff2}(e) we plot $\delta E_{\textrm{qp}}$ for the same value of $\tilde{J}=0.5t$. Close to the region $x_{12}\approx 0$, we do not observe pronounced oscillations in the magnetic configuration along the constant $x_{12}$ lines, similar to that in Fig.~\ref{fig:ediff2}(b) for weak $\tilde{J}$.  As the magnitude of $x_{12}$ increases, the magnetic configuration starts oscillating between the FM and the AFM phases as shown in the figure. However, the number of constant $x_{12}$ lines along which the oscillations occur is much less compared to that in Fig.~\ref{fig:ediff2}(b). This behaviour is consistent with the features seen in Fig. \ref{fig:ediff}(b), where we observe less  oscillations between different magnetic configurations in $\delta E_{\textrm{qp}}$ for such a strong value of $\tilde{J}$. 

Finally, Fig.~\ref{fig:ediff2}(f) shows the total ground state energy difference $\delta E_{\textrm{g}}$, which  is a sum of the YSR and the quasiparticle contributions as also seen previously. Similar to the features in Fig.~\ref{fig:ediff2}(c), the phase diagram of $\delta E_{\textrm{g}}$ is dominated by the bulk contributions for $|x_{12}|\lesssim \lambda_F$ and thus does not exhibit oscillations in the magnetic configuration in a region around $x_{12}=0$. Beyond $x_{12}=\lambda_F$, the YSR contribution wins over the bulk one and, as a result, the phase diagram of $\delta E_{\textrm{g}}$ is similar to the one  of $\delta E_{\textrm{YSR}}$ in Fig.~\ref{fig:ediff2}(d).

To conclude, we increase $\tilde{J}$ further and in the last row of Fig.~\ref{fig:ediff2}, we have $\tilde{J}=1.5 t$.  At such a large value of the exchange interaction, the YSR states again move towards the superconducting gap edge. The boundary-induced hybridization gives rise to oscillations in the magnetic configuration between the FM and the AFM phases as seen in Fig. \ref{fig:ediff2}(g).

In  Fig.~\ref{fig:ediff2}(h) obtained from the quasiparticle energies, there is no oscillation along the constant $x_{12}$ lines close to the region $x_{12}\approx 0$, as also seen above. With increasing $x_{12}$, we find that the  oscillations between the two magnetic configurations occur only along the lines satisfying the condition $2x_{12}k_F=(2n+1)\pi/2$. The phase diagram for $\delta E_{\textrm{g}}$ in Fig.~\ref{fig:ediff2}(i) gets contributions from both the YSR and the quasiparticle states as seen above for two other values of $\tilde{J}$. The magnetic ground state follows the quasiparticle states for $x_{12}$ smaller than $\lambda_F$, and beyond that $\delta E_{\textrm{g}}$ is dominated by the YSR contribution as shown in Fig.~\ref{fig:ediff2}(i).

\section{Conclusions}

We investigated the effects of a boundary on the YSR states and on the magnetic ground state of two classical spins in a 1D superconductor. We showed the change in the hybridization between the YSR states as the impurities move close to the boundary. For small  exchange interaction strength (between impurity spin and electron spins) compared to the Fermi energy, we calculated the RKKY interaction between the magnetic impurities in a semi-infinite system. The RKKY interaction, not only depends on the inter-impurity distance, but also on the distances of the impurities from the boundary. It is therefore possible to drive a phase transition between different magnetic ground state configurations by solely changing the impurity-boundary distances. While it is expected that the boundary will induce Friedel oscillations in the wavefunctions, the possibility that this leads to a phase transition is rather
surprising and interesting.
We also found that depending on the distance between the magnetic impurities, the boundary effect can be suppressed or enhanced. Thus, the inter-impurity distance acts as a tuning parameter of the boundary-induced physics. Next, we numerically explored the boundary effect for small exchange interactions and away from this limit. Our numerical plots exhibit phase transitions occurring as a function of the impurity-boundary distances, similar to the analytical results for weak coupling.  Moreover, the distinctive features of the boundary effects, dependent on the choice of the inter-impurity distance, also remain unaltered in the limit of strong exchange interactions. We observe that the numerically obtained phase diagram demonstrates that the energy difference between the FM and the AFM ground state is governed by the sum of both the YSR and  the quasiparticle contributions. Our findings thus re-emphasize the importance of including both the YSR bound states and the quasiparticle states when determining the total magnetic ground state. It is straightforward to generalize our results to account for spin-orbit interactions in a 1D system as it can be easily absorbed into the tilt of one of the spin impurities by making use of the position-dependent gauge\cite{bruno}, resulting in magnetic configurations that are not collinear.  Finally, we note that the predicted behavior of the YSR states can be also observed in semiconducting nanowires with proximity-induced superconductivity, where the presence of the YSR states was demonstrated in recent experiments\cite{prada,s1,s2,s3,s4,s5}.

\section{acknowledgements}

We acknowledge support from the Swiss National Science Foundation and NCCR QSIT. This project received funding from the European Union's Horizon 2020 research and innovation program (ERC Starting Grant, grant agreement No 757725). SH was also supported by the Center for Molecular Magnetic Quantum Materials, 
an Energy Frontier Research Center funded by the U.S. Department of Energy, Office of Science, Basic Energy Sciences under Award No. DE-SC0019330.


\begin{thebibliography} {99}


\bib{1} A. I. Rusinov, Sov. Phys. JETP {\bf 29}, 1101 (1969).

\bib{2} L. Yu, Acta Phys. Sin. {\bf 21}, 75 (1965).

\bib{3} H. Shiba, Prog. Theor. Phys. {\bf 40}, 435 (1968).

\bib{4} P. Schlottmann, Phys. Rev. B {\bf 13}, 1 (1976).

\bib{5} A. I. Rusinov, Zh. Eksp. Teor. Fiz., Pis’ma Red. {\bf 9}, 146 (1968) [JETP Lett. {\bf 9}, 85 (1969)].

\bib{6} A. Sakurai, Prog. Theor. Phys.{\bf 44}, 1472 (1970).

\bib{baur} W. Bauriedl, P.  Ziemann, and W. Buckel, Phys. Rev. Lett. {\bf 47}, 1163 (1981).

\bib{8} M. E. Flatt\'{e} and J. M. Byers, Phys. Rev. Lett. {\bf 78}, 3761 (1997).

\bib{9} M. I. Salkola, A. V. Balatsky, and J. R. Schrieffer, Phys. Rev. B {\bf 55}, 12648 (1997).

\bib{10} M. E. Flatt\' e and D. E. Reynolds, Phys. Rev. B {\bf 61}, 14810 (2000).

\bib{11} D. K. Morr and J. Yoon, Phys. Rev. B {\bf 73}, 224511 (2006).

\bib{12} A. V. Balatsky, I. Vekhter, and J.-X. Zhu, Rev. Mod. Phys. {\bf 78}, 373 (2006).

\bib{t4a} C. P. Moca, E. Demler, B. Janka, and G. Zarand, Phys. Rev. B, {\bf 77}, 174516 (2008).

\bib{14} N. Y. Yao, L. I. Glazman, E. A. Demler, M. D. Lukin, and J. D. Sau, Phys. Rev. Lett. {\bf 113}, 087202 (2014).

\bib{15} N. Y. Yao, C. P. Moca, I. Weymann, J. D. Sau, M. D. Lukin, E. A. Demler, and G. Zarand, Phys. Rev. B {\bf 90}, 241108(R) (2014).

\bib{16} A. A. Zyuzin and D. Loss, Phys. Rev. B {\bf 90}, 125443 (2014).

\bib{17} S. Hoffman, J. Klinovaja, T. Meng, and D. Loss,  Phys. Rev. B {\bf 92}, 125422 (2015).

\bib{meng} T. Meng, J. Klinovaja, S. Hoffman, P. Simon, and D. Loss, Phys. Rev. B {\bf 92}, 064503 (2015).

\bib{pascal} V. Kaladzhyan, C. Bena, and P. Simon, Phys. Rev. B {\bf 93}, 214514 (2016). 

\bib{vard}  V. Kaladzhyan, S. Hoffman, and M. Trif, Phys. Rev. B {\bf 95}, 195403 (2017).

\bib{t1}  A. Ptok, S. Głodzik, and T. Domański, Phys. Rev. B {\bf 96}, 184425 (2017).

\bib{tra}  S. K\"orber, B. Trauzettel, and O. Kashuba, Phys. Rev. B {\bf 97} 184503 (2018).

\bib{trif} A.  Mishra, S. Takei, P. Simon, and M. Trif,  arXiv:2007.15392.



\bib{7}A. Yazdani, B. A. Jones, C. P. Litz, M. F. Crommie, and D. M. Eigler, Science {\bf 275}, 1767 (1997).

\bib{exp11} A. Yazdani, C. M. Howald, C. P. Lutz, A. Kapitulnik, and D. M. Eigler, Phys. Rev. Lett. {\bf 83}, 176 (1999).

\bib{13} S. -H. Ji, T. Zhang, Y. -S. Fu, X. Chen, X. -C. Ma, J. Li, W. -H. Duan, J. -F. Jia, and Q. -K. Xue, Phys. Rev. Lett. {\bf 100}, 226801 (2008).

\bib{hatter} N. Hatter, B. W. Heinrich, M. Ruby, J. I. Pascual, K. J. Franke, Nature Communications {\bf 6},  8988 (2015).

\bib{ruby} M. Ruby, F. Pientka, Y. Peng, F. von Oppen, B. W. Heinrich, and K. J. Franke, Phys. Rev. Lett. {\bf 115}, 087001 (2015). 

\bib{menard} G. C. Menard, S. Guissart, C. Brun, S. Pons, V. S. Stolyarov, F. Debontridder, M. V. Leclerc, E. Janod, L. Cario, D. Roditchev, P. Simon, and T. Cren,  Nature Physics {\bf 11}, 1013 (2015).

\bib{exp4} M. Ruby, Y. Peng, F. von Oppen, B.W. Heinrich, K.J. Franke, Phys. Rev. Lett. {\bf 117}, 186801 (2016).

\bib{exp5a} A. Jellinggaard, K. Grove-Rasmussen, M. H. Madsen, J. Nygård,  Phys. Rev. B {\bf 94}, 064520 (2016).

\bib{exp5}  D.-J. Choi, C. Rubio-Verda, J. de Bruijckere, M. M. Ugeda, N. Lorente, J. I. Pascual,  Nat. Comm., {\bf 8} 15175 (2017).

\bib{exp6} B. W. Heinrich, J. I. Pascual, and K. J. Franke, Progress in Surface Science {\bf 93}, 1, (2018).

\bib{exp2} L. Farinacci, G. Ahmadi, G.Reecht, M. Ruby, N. Bogdanoff, O. Peters, B. W. Heinrich, F. von Oppen, and K.  J. Franke, Phys. Rev. Lett. {\bf 121}, 196803 (2018).

\bib{exp7} S. Kezilebieke, M. Dvorak, T. Ojanen, and P. Liljeroth, Nano Lett.  {\bf 18}, 2311 (2018).

\bib{exp3}  V. Perrin, F. L. N. Santos, G. C. Menard, C. Brun, T. Cren, M. Civelli, and P.  Simon, Phys. Rev. Lett. {\bf 125}, 117003 (2020).

\bib{exp8} A. Kamlapure,  L. Cornils, J. Wiebe, and R.  Wiesendanger, Nat. Commun. {\bf 9}, 3253 (2018).



\bib{s2} A. Jellinggaard, K.Grove-Rasmussen, M. Hannibal Madsen, and J. Nygard, Phys. Rev. B {\bf 94}, 064520 (2016).

\bib{s3}  J. O. Island, R. Gaudenzi, J. de Bruijckere, E. Burzuri, C. Franco, M. Mas-Torrent, C. Rovira, J. Veciana, T. M. Klapwijk, R. Aguado, and H. S. J. van der Zant
Phys. Rev. Lett. {\bf 118}, 117001 (2017).

\bib{s1} K. Grove-Rasmussen, G. Steffensen, A. Jellinggaard, M. H. Madsen, R. Zitko, J. Paaske, and  J. Nygard, Nature Communications {\bf 9},  2376 (2018).

\bib{prada} E. Prada, P. San-Jose, M. W. A. de Moor, A. Geresdi, E. J. H. Lee, J. Klinovaja, D. Loss, J. Nygard, R. Aguado, and L. P. Kouwenhoven, 
Nature Reviews Physics {\bf 2}, 575 (2020).

\bib{s4} J. C. Estrada Saldana, A. Vekris, R. Zitko, G. Steffensen, P. Krogstrup, J. Paaske, K. Grove-Rasmussen, and J. Nygard, Phys. Rev. B {\bf 102}, 195143 (2020).

\bib{s5}  M. Valentini, F. Penaranda, A. Hofmann, M. Brauns, R. Hauschild, P.  Krogstrup, P. San-Jose, E. Prada, R. Aguado, G. Katsaros, arXiv:2008.02348.



\bib{19} J. Klinovaja, P. Stano, A. Yazdani, and D. Loss, Phys. Rev. Lett. {\bf 111}, 186805 (2013).

\bib{20} M. M. Vazifeh and M. Franz, Phys. Rev. Lett. {\bf 111}, 206802 (2013).

\bib{21} B. Braunecker and P. Simon, Phys. Rev. Lett. {\bf 111}, 147202 (2013).

\bib{18} S. Nadj-Perge, I. K. Drozdov, B. A. Bernevig, and A. Yazdani, Phys. Rev. B {\bf 88}, 020407(R) (2013).

\bib{22} S. Nakosai, Y. Tanaka, and N. Nagaosa, Phys. Rev. B {\bf 88}, 180503 (2013).

\bib{23} F. Pientka, L. I. Glazman, and F. von Oppen, Phys. Rev. B {\bf 88}, 155420 (2013).

\bib{24} K. P\"oyh\"onen, A. Weststr\"om, J. R\"ontynen, and T. Ojanen, Phys. Rev. B {\bf 89}, 115109 (2014).

\bib{25} I. Reis, D. J. J. Marchand, and M. Franz, Phys. Rev. B {\bf 90}, 085124 (2014).

\bib{kotetes} A. Heimes, D. Mendler, and P.  Kotetes, New J. Phys. {\bf 17}, 023051 (2015).

\bib{sh} S. Hoffman, J. Klinovaja, and D. Loss, Phys. Rev. B {\bf 93}, 165418 (2016).

\bib{ando} G. M. Andolina and P. Simon, Phys. Rev. B {\bf 96}, 235411 (2017). 

\bib{t4}  K. Bj\"ornson, A. V. Balatsky, and A. M. Black-Schaffer, Phys. Rev. B {\bf 95}, 104521  (2017).

\bib{awoga} O.A Awoga and A. M. Black-Schaffer, Phys. Rev. B {\bf 97}, 214515 (2018).

\bib{rev2} D.-J.  Choi, et al., Rev. Mod. Phys. {\bf 91}, 041001 (2019).

\bib{rev3} R. Pawlak, S. Hoffman, J. Klinovaja, D. Loss, and E. Meyer, Progress in Particle and Nuclear Physics {\bf 107}, 1 (2019).

\bib{thriler}  A. Theiler, K. Bj\"ornson, and A. M. Black-Schaffer, Phys. Rev. B {\bf 100}, 214504 (2019).

\bib{black} M. Mashkoori and A. Black-Schaffer, Phys. Rev. B {\bf 99}, 024505 (2019).





\bib{26} S. Nadj-Perge, I. K. Drozdov, J. Li, H. Chen, S. Jeon, J. Seo, A. H. MacDonald, B. A. Bernevig, and A. Yazdani, Science {\bf 346}, 602 (2014).

\bib{r1} M. Ruby, F. Pientka, Y. Peng, F. von Oppen, B. W. Heinrich, and K. J. Franke, Phys. Rev. Lett. 115, 197204 (2015).

\bib{27} R. Pawlak, M. Kisiel, J. Klinovaja, T. Meier, S. Kawai, T.Glatzel, D. Loss, and E. Meyer, npj Quantum information {\bf 2}, 16035 (2016).

\bib{feld} B. E. Feldman, M. T. Randeria, J. Li, S. Jeon, Y. Xie, Z. Wang, I. K. Drozdov, B. A. Bernevig, and A. Yazdani, Nature Physics {\bf 13}, 286 (2017).

\bib{w1} H. Kim,  et al.,  Sci. Adv. {\bf 4}, eaar5251 (2018).





\bib{28a} M. A. Ruderman and C. Kittel, Phys. Rev. {\bf 96}, 99 (1954).

\bib{28b}  T. Kasuya, Prog. Theor. Phys. {\bf 16}, 45 (1956).

\bib{28c}  K. Yosida, Phys. Rev. {\bf 106}, 893 (1957).

\bib{giulani} G. Giuliani and G. Vignale, {\it Quantum Theory of the Electron Liquid} (Cambridge University Press, Cambridge, UK, 2005).

\bib{bruno2} P. Bruno, Phys. Rev. B {\bf 52}, 411 (1995).

\bib{sch} N. F. Schwabe et al., Phys. Rev. B {\bf 54}, 12953 (1996).

\bib{egger} R. Egger and H. Schoeller, Phys. Rev. B {\bf 54}, 16337 (1996).

\bib{bruno}  H. Imamura, P. Bruno, and Y.  Utsumi, Phys. Rev. B {\bf 69}, 121303(R) (2004).

\bib{29} P. Simon and D. Loss, Phys. Rev. Lett. {\bf 98}, 156401 (2007).

\bib{30} B. Braunecker, P. Simon, and D. Loss, Phys. Rev. Lett. {\bf 102}, 116403 (2009).

\bib{schaffer} A. M. Black-Schaffer, Phys. Rev. B {\bf 81}, 205416 (2010). 

\bib{chesi} S. Chesi and D. Loss, Phys. Rev. B {\bf 82}, 165303 (2010). 

\bib{31} B. Braunecker, G. I. Japaridze, J. Klinovaja, and D. Loss, Phys. Rev. B {\bf 82}, 045127 (2010).

\bib{kogan} E. Kogan, Phys. Rev. B {\bf 84}, 115119 (2011). 

\bib{klinovaja} J. Klinovaja, and D. Loss, Phys. Rev. B {\bf 87}, 045422 (2013). 

\bib{32} T. Meng, P. Stano, J. Klinovaja, and D. Loss, Eur. Phys. J. B {\bf 87}, 203 (2014).

\bib{hsu1} C.-H. Hsu, P. Stano, J. Klinovaja, and D. Loss, Phys. Rev B {\bf 92}, 235435 (2015). 

\bib{t5} M. V. Hosseini and M. Askari, Phys. Rev. B {\bf 92}, 224435 (2015).

\bib{flensberg} M. Schecter, M. S. Rudner, and K. Flensberg, Phys. Rev. Lett. {\bf 114}, 247205 (2015). 

\bib{hsu2} C.-H. Hsu, P. Stano, J. Klinovaja, and D. Loss, Phys. Rev. B {\bf 97}, 125432 (2018).

\bib{t10} O. M. Yevtushenko and V. I. Yudson, Phys. Rev. Lett. {\bf 120}, 147201 (2018).

\bib{t6} V. Kaladzhyan, A. A. Zyuzin, and P. Simon, Phys. Rev. B {\bf 99}, 165302 (2019).

\bib{t7}  G. C. Paul, S. K. Firoz Islam, and A. Saha, Phys. Rev. B {\bf 99}, 155418 (2019).

\bibitem{henry} H. F. Legg and B. Braunecker, Scientific Reports {\bf 9}, 17697 (2019).

\bib{t8} E. Kogan, C {\bf 5} 14 (2019).

\bib{t9}  A. M. Tsvelik and O. M. Yevtushenko, Phys. Rev. B {\bf 100}, 165110 (2019).

\bib{st} P. Stano, J. Klinovaja, A. Yacoby, and D. Loss, Phys. Rev. B {\bf 88}, 045441, (2013).

\bib{36} Y. Yafet, Phys. Rev. B {\bf 36}, 3948 (1987).

\bib{37} V. I. Litvinov and V. K. Dugaev, Phys. Rev. B {\bf 58}, 3584
(1998).

\bib{abri} A. A. Abrikosov, Fundamentals of the Theory of Metals, Vol. 1 (Elsevier, Amsterdam, 1988).

\bib{aristov} D. Aristov, S. Maleyev, and A. Yashenkin, Z. Phys. B {\bf 102}, 467 (1997).

\bib{larkin} V. M. Galitski and A. I. Larkin, Phys. Rev. B {\bf 66}, 064526 (2002).


\end{thebibliography}
\end{document}